\documentclass[a4paper,11pt]{article}
\usepackage[left=2cm,right=2cm,top=2cm,bottom=2cm]{geometry}
\usepackage[usenames,dvipsnames,svgnames,table]{xcolor}
\usepackage{psfrag,amsmath,amsfonts,amssymb,latexsym,amsthm,verbatim,color,lscape,multirow,graphicx,subcaption}
\usepackage[figuresright]{rotating}
\usepackage[compress]{natbib}
\usepackage{url}
\usepackage{hyperref}
\hypersetup{colorlinks,
citecolor=black,
filecolor=black,
linkcolor=black,
urlcolor=black,
pdftex}

\usepackage{tikzducks}
\usepackage{booktabs}
\usepackage[displaymath, mathlines,pagewise]{lineno}
\usepackage{colortbl}
\definecolor{gray}{rgb}{0.9,0.9,0.9}

\usepackage{times}
\usepackage{blindtext}
\usepackage[font={footnotesize,it}]{caption}

\usepackage{sectsty}  
\subsectionfont{\normalsize\bf}
\sectionfont{\large\bf}

\usepackage{tikz}
\usetikzlibrary{shapes.geometric}

\newcommand{\blackcircle}{\raisebox{0pt}{\tikz{ 
\node[circle,fill,black,draw,scale=.5pt,rotate=0] {};\draw[-,black,solid,line width = 1.0pt](-4mm,.3pt) -- (4mm,.3pt)}}}

\newcommand{\greendiamond}{\raisebox{0pt}{\tikz{ 
\node[rectangle,fill,black!20!green,draw,scale=.5pt,rotate=45] {};\draw[-,black!20!green, dashdotted,line width = 1.0pt](-4.2mm,.3pt) -- (4.2mm,.3pt)}}}

\newcommand{\violetrectangle}{\raisebox{0pt}{\tikz{ 
\node[rectangle,fill,black!30!violet,draw,scale=.7pt,rotate=0] {};\draw[-,black!30!violet,densely dashed,line width = 1.0pt](-4mm,.3pt) -- (4mm,.3pt)}}}

\definecolor{deepcerise}{rgb}{0.85, 0.2, 0.53}  
\newcommand{\pinktriangle}{\raisebox{0.3pt}{\tikz{ 
\node[isosceles triangle, isosceles triangle apex angle=60,fill,deepcerise,draw,scale=.4pt,rotate=90] {};\draw[-,deepcerise, dotted,line width = 1.0pt](-4mm,1pt) -- (4mm,1pt)}}}

\def\p{\partial}

\def\R{\mathbf{R}}
\def\y{\mathbf{y}}
\def\Y{\mathbf{Y}}

\def\x{\mathbf{x}}

\def\X{\mathbf{X}}

\def\th{\theta}
\def\thbf{\boldsymbol{\theta}}

\def\R{\mathbf{R}}
\def\y{\mathbf{y}}
\def\Y{\mathbf{Y}}

\def\th{\theta}
\def\thbf{\boldsymbol{\theta}}

\long\def\symbolfootnote[#1]#2{\begingroup
\def\thefootnote{\fnsymbol{footnote}}\footnote[#1]{#2}\endgroup}

\let\proglang=\textsf
\newcommand{\pkg}[1]{{\fontseries{b}\selectfont #1}}

\newcommand{\tr}{\mathrm{tr}} 

\usepackage[normalem]{ulem}

\begin{document}

\title{Factor tree copula models for item response data}

\author{Sayed H. Kadhem  \and Aristidis~K.~Nikoloulopoulos \footnote{Correspondence to: \href{mailto:a.nikoloulopoulos@uea.ac.uk}{a.nikoloulopoulos@uea.ac.uk}, Aristidis K. Nikoloulopoulos, School of Computing Sciences, University of East Anglia, Norwich NR4 7TJ, U.K.} }

\date{}

\maketitle

\begin{abstract}
\baselineskip=24pt
\noindent
Factor copula models  for item response data are more interpretable and fit better than (truncated) vine copula models when dependence can be explained through latent variables, but are not robust to violations of conditional independence. To circumvent these issues, truncated vines and factor copula models for item response data are  joined to
define a combined model, the so called factor tree copula model,  with individual benefits from each of the two approaches. Rather than adding factors and causing computational problems and difficulties in interpretation and identification, a truncated vine structure is assumed on the residuals conditional on one or two latent variables. This structure can be better explained as a conditional dependence given a few interpretable latent variables. On the one hand the parsimonious feature of factor models remains intact and   any residual dependencies are being taken into account on the other.
We discuss estimation along with model selection. In particular we propose model selection algorithms to choose a plausible factor tree copula model to capture the (residual) dependencies among the item responses. 
Our general methodology is demonstrated with an extensive simulation study and illustrated by analysing   Post Traumatic Stress Disorder.  

\noindent \textbf{Key Words:}   Conditional dependence; Factor copula models; Markov trees; Spanning tree algorithm; Truncated vine copula models.
\end{abstract}

\baselineskip=24pt

\section{Introduction}
Factor or conditional independence models are widely used techniques for analysing item response data using much fewer unobserved/latent variables or factors \citep{Bartholomew&Knott&Moustaki2011-Wiley}.  These are natural if the dependence amongst the $d$ observed variables or items is assumed to arise from $p$ latent variables with $p<<d$. They are parsimonious models and favourable for large dimensions as the number of parameters is $\mathcal{O}(d)$ instead of $\mathcal{O}(d^2)$.  Nevertheless, factor models mainly  assume that the items  are conditionally independent given some latent variables. This assumption implies that the dependence amongst the observed variables is fully accounted for by the factors with no remaining dependence. This could lead to biased estimates if the strict assumption of conditional independence is violated \citep{Braeken-etal2007,Sireci-etal1991,Chen&Thissen1997,Yen1993}. The conditional independence assumption is violated if there exists  local or residual dependence. 
Mitigating the residual dependence might be achieved by adding more latent variables to the factor model, but at the expense of computational problems and difficulties in interpretation and identification.

To circumvent these problems,  the items can be allowed to interrelate by forming a dependence structure with conditional dependence given a few interpretable
latent variables.
 In this way, on the one hand the parsimonious feature of factor models remains intact and   any residual dependencies are being taken into account on the other. This can be achieved by incorporating copulas into the conditional distribution of factor models in order  to provide a conditional dependence structure given very few latent variables.  
Such copula approaches for item response data are  proposed by \cite{Braeken-etal2007,Braeken-etal2013} and \cite{Braeken2011} who explored the use of Archimedean copulas or a mixture of the independence and comonotonicity copulas to capture the residual dependence of  traditional item response theory models. Therein simple copulas have been used for subgroups of items that are chosen from the context with homogeneous within-subgroup dependence.  This is due to the fact that Archimedean copulas  allow only for exchangeable dependence with  a narrower range as the dimension increases \citep{mcneil&neslehova08}.

Without a priori knowledge of obvious subgroups of items that are approximately exchangeable, we will propose a more general residual dependence approach that makes use of truncated regular vine copula models \citep{Brechmann-etal2012}. 
 Within a vine copula specification, no such
restrictions need to be made.
Regular vine copulas are a flexible class of models that are constructed from a set of bivariate copulas in hierarchies or tree levels \citep{Joe1996,Bedford&Cooke2001,Bedford&Cooke2002,Kurowicka&Cooke2006,Kurowicka&Joe2011,Joe2014-CH}. 
A $d$-dimensional regular vine copula
can cover flexible dependence structures, different from assuming simple linear correlation structures, tail independence and normality \citep{Nikoloulopoulos&Joe&Li2012-CSDA}, through the specification
of $d-1$ bivariate  parametric copulas 
at level 1 and $\binom{d-1}{2}$
bivariate conditional parametric copulas at higher levels; at level $\ell$  for
$\ell=2,\ldots,d-1$, there are $d-\ell$ bivariate conditional copulas
that condition on $\ell-1$ variables. 
\cite{Joe&Li&Nikoloulopoulos2010} have shown that in order for a vine copula to have (tail) dependence for all bivariate margins, it is only necessary for the bivariate copulas in level 1 to have (tail) dependence and it is not necessary for the conditional bivariate copulas in levels $2,\ldots,d-1$ to have (tail) dependence. That provides 
the  theoretical justification for the idea to model the  dependence in the first level and then just use the independence copulas to model conditional dependence at higher levels without sacrificing the tail dependence of the vine copula distribution. That is 
the 1-truncated vine copula has $d-1$ parametric bivariate copulas in the 1st level of the vine and independence copulas in all the remaining levels of the vine (truncated after the 1st level).
This truncation, as per the terminology in \citep{Brechmann-etal2012}, provides  a parsimonious vine copula model.   
The 1-truncated vine copula can provide, 
with appropriately chosen linking copulas,
 asymmetric dependence structure as well as tail dependence (dependence among extreme values). \cite{Joe&Li&Nikoloulopoulos2010} have shown that  by choosing bivariate linking copulas appropriately, vine copulas  can have a flexible range of lower/upper tail dependence and different lower/upper tail dependence parameters for each bivariate margin. Choices of copulas with upper or lower tail dependence are better if the items have more joint upper or lower tail probability than would be expected with the discretized multivariate normal (MVN) model \citep{Muthen1978-PKA}. 
Note in passing that the discretized MVN distribution is a special case of the vine copula model with discrete margins. If all bivariate copulas are bivariate normal (BVN) in the vine copula model, then the resulting model is the discretized MVN. 

To define the conditional independence part of the model we also use truncated vine copulas rather than the traditional factor models for item response  in \cite{Braeken-etal2007,Braeken-etal2013} and \cite{Braeken2011}. 
\cite{Nikoloulopoulos2015-PKA} have proposed factor copula models for item response data. These factor models can be explained as truncated canonical vines rooted at the latent variables. The canonical vine is a boundary case of regular vine copulas, which is suitable if there exists a (latent) variable that drives the dependence among the items.  For the first factor there are  bivariate copulas that couple each item to the first latent variable and for the second factor there are copulas that link each item to the second latent variable conditioned on the first factor (leading to conditional dependence parameters), etc. Factor copula models  with appropriately chosen linking copulas will be useful when the items (a) have more probability in joint upper or lower tail than would be expected with a discretized multivariate normal, or (b)  can be considered as discretized maxima/minima or mixtures of discretized means rather than discretized means \citep{Nikoloulopoulos2015-PKA}.

The proposed parsimonious approach, that requires no priori knowledge of the subgroups of items,   can be explained as a truncated regular vine copula model that involves both observed and latent variables; but, more simply, we derive the models as conditional dependence models  with a few interpretable latent variables that model the residual dependence of the factor copula model via an 1-truncated vine copula. 
The factor copula model explains most of the dependence and the remaining dependence can be further accounted for  by an 1-truncated vine copula conditioned on the factors. 
\cite{Brechmann2014} and \cite{Joe2018} initiated the study of such conditional dependence models with a unidimensional factor/latent variable for continuous data. The combined 1-factor and 1-truncated vine model for continuous data in \cite{Brechmann2014}  is restricted to Gaussian dependence, but  \cite{Joe2018} proposed a 
combination of an 1-factor copula model with 1-truncated vine copula model   with non-Gaussian  bivariate  copulas.  Our models for item response are discrete counterparts  of the models in \cite{Brechmann2014} and \cite{Joe2018} with interpretation and technical details that are quite different and provide an extension to more than one factors.

The remainder  of the paper proceeds as follows. In Section \ref{sec:Fvine}, we introduce the combined factor/truncated  vine copula models for item response data. Section \ref{sec:Estimation} provides estimation techniques and computational details. Section \ref{sec:ModelSelection}  discusses  vine tree and bivariate copula selection. Section \ref{sec:Simulations} has an extensive simulation study to assess the estimation techniques and model selection algorithms. Our methodology is illustrated using real data in Section \ref{sec:Applications}. We conclude with some discussion in Section  \ref{sec:Discussion}, followed by a brief section with software details.

\section{Factor tree copula models for item response} \label{sec:Fvine}
This section introduces the theory of the combined factor/truncated vine copula models for item response data. 
Before that, the first two sections provide some background about factor \citep{Nikoloulopoulos2015-PKA} and truncated vine \citep{Panagiotelis&Czado&Joe2012,Panagiotelis-etal-2017-CSDA} copula  models for discrete  responses.

\subsection{Factor copula models}\label{Sec:FactorCopula}
We first introduce the notation used in this paper. 
Let $\Y=\{Y_{1}, \ldots,Y_{d}\}$ denote the vector with the  item response variables that are all measured on an ordinal scale; $Y_{j}\in \{0,\ldots,K_{j}-1\}$. Let the cutpoints in the uniform $U(0,1)$ scale for the $j$th item be $a_{j,k}$, $k=1,\ldots,K-1$,  with $a_{j,0}=0$ and $a_{j,K}=1$. These correspond to $a_{j,k}=\Phi(\alpha_{j,k})$, where $\alpha_{j,k}$ are cutpoints in the normal $N(0, 1)$ scale.

The $p$-factor model assumes that  $\Y$, with corresponding realizations $\y=\{y_{1}, \ldots,y_{d}\}$, is conditionally
independent given the $p$-dimensional latent vector $\X=(X_1,\ldots,X_p)$. The joint probability mass function (pmf) of the $p$-factor model is 
\begin{equation}\label{p-factor-pmf}
\pi_d(\mathbf{y})=\Pr(Y_1=y_1,\ldots,Y_d=y_d)= \int\prod_{j=1}^d\Pr(Y_j=y_j|X_1=x_1,
 \ldots,X_p=x_p)\, dF_{\X}(x),
\end{equation}
where $F_{\X}$ is the distribution of the latent vector $\X$. 
The factor copula methodology uses a set of bivariate copulas that link the items to  the latent variables to specify
$\Pr(Y_j=y_j|X_1=x_1, \ldots,X_p=x_p)$. Below we include the theory for one and two factors.

For the 1-factor model, let $X_1$ be a latent variable that is standard uniform.
From \cite{Sklar1959}, there is a
bivariate copula $C_{X_1j}$
such that $\Pr(X_1\le x, Y_j\le y)=C_{X_1j}\bigl(x,F_j(y)\bigr)$ for $0\leq x\leq 1$ where $F_j(y)=a_{j,y+1}$ is the
cdf of $Y_j$. Then it follows that 
\begin{equation}
  F_{j|X_1}(y|x):=\Pr(Y_j\le y|X_1=x) = {\p C_{X_1j}(x,a_{j,y+1})\over\p x}=C_{j|X_1}(a_{j,y+1}|x).
  \label{eq-condjL1}
\end{equation}

Hence, the pmf for the 1-factor copula  model becomes
\begin{eqnarray*}
\label{1-pdf}
\pi_d(\mathbf{y})&=&\int_0^1\prod_{j=1}^d\Pr(Y_j=y_j|X_1=x)\,dx
=\int_0^1\prod_{j=1}^d f_{j|X_1}(y_j|x) \,dx, 
\end{eqnarray*}
where 
\begin{equation}\label{1-factor-inner}
f_{j|X_1}(y|x)=
C_{j|X_1}(a_{j,y+1}|x) -  C_{j|X_1}(a_{j,y}|x).
\end{equation}

For the 2-factor copula model, let $X_1,X_2$ be latent variables that are independent uniform $U(0,1)$ random variables.
Let $C_{X_1j}$ be defined as in the 1-factor copula model
and  $C_{X_2j}$ be a bivariate copula such that 
$$
\Pr(X_2\le x_2,Y_j\le y|X_1=x_1) =C_{X_2j}\bigl(x_2,F_{j|X_1}(y|x_1)\bigr),
$$
where $F_{j|X_1}$ is given in (\ref{eq-condjL1}). Then for $0\leq x_1,x_2\leq1$, 
\begin{align}
F_{X_2j|X_1}(x_2,y|x_1):=\Pr(Y_j &  \le   y|X_1=x_1,X_2= x_2)
= {\p\over \p x_2} \Pr(X_2\le x_2,Y_j\le y|X_1=x_1) \nonumber  \\
&={\p\over \p x_2} C_{X_2j}\Bigl(x_2,F_{j|X_1}(y|x_1)\Bigr)=
   C_{j|X_2}\Bigl(F_{j|X_1}(y|x_1)|x_2\Bigr).
    \label{eq-condjL2}
\end{align}

Hence, the pmf for the 2-factor copula model is 
\begin{eqnarray*}
\label{2-pdf}
\pi_d(\mathbf{y})&=&
\int_0^1\int_0^1\prod_{j=1}^d\Pr(Y_j=y_j|X_1=x_1,X_2=x_2)\,dx_1 dx_2 \nonumber\\
&=&\int_0^1\int_0^1\prod_{j=1}^d
f_{X_2j|X_1}\bigl(x_2,y_j|x_1\bigr)\,dx_1 dx_2, 
\end{eqnarray*}
where 
\begin{equation}\label{2-factor-inner}
f_{X_2j|X_1}(x_2,y|x_1)=
C_{j|X_2}\Bigl(F_{j|X_1}(y|x_1)|x_2\Bigr)-C_{j|X_2}\Bigl(F_{j|X_1}(y-1|x_1)|x_2\Bigr).
\end{equation}

\subsection{1-truncated vine copula models} 
\label{Sec:VineCopula}

Vine copula models are flexible tools to analyse dependence structures and have been popular in many application areas  \citep{Kurowicka&Joe2011}. They involve $d-1$ trees,  the first tree represents dependence (as edges) amongst $d$ variables (as nodes). Then the edges become nodes in the next tree, involving the conditional dependencies given a common variable. This process continues until tree $d - 1$ that includes two nodes and one edge, representing conditional dependence of two variables given $d - 2$ variables \citep{Chang&Joe2019}.

If one is restricted to the first tree, that is truncation at level 1, then the result is a Markov tree
dependence structure where two variables not connected by an edge are conditionally independent given the variables in
the tree between them.
In a Markov tree or 1-truncated vine with  $d$ variables, $d-1$
of the $d(d - 1)/2$ possible pairs are identified as the edges of a tree with $d$ nodes
corresponding to the items, i.e., there are a total of $d-1$ edges, where  two connected pairs of items form an edge.
Let $j$ and $k$ be indices for any pairs of items with $1 \leq k < j \leq d$. For a given vine tree structure, let $\mathcal{E}$ denote the set of edges. Each edge of $jk \in \mathcal{E}$  
is represented with a bivariate copula $C_{jk}$ such that
$$\Pr(Y_j\leq y_j,Y_k \leq y_k)=C_{jk}\bigl(F_j(y_j),F_k(y_k)\bigr)=C_{jk}(a_{j,y_j+1},a_{k,y_k+1}).$$

Since the densities of  vine copulas can be factorized in terms of bivariate linking copulas and lower-dimensional margins, they are computationally tractable for high-dimensional continuous variables. Nevertheless,
the cdf of $d$-dimensional vine copula lacks a closed form and requires  $(d-1)$-dimensional integration \citep{Joe1997-CH}. Hence, in order to derive the $d$-dimensional  pmf using  finite differences of the $d$-dimensional cdf (e.g., \citealt{Braeken-etal2007} or \citealt{Nikoloulopoulos2013a}) poses non-negligible numerical challenges.
This problem has been solved by   \cite{Panagiotelis&Czado&Joe2012} who decomposed the $d$-dimensional pmf  into finite differences of bivariate copula cdfs. Hence, the pmf of an 1-truncated  vine model  takes the form 
\begin{equation}\label{vine-pmf}
\pi_d(\y)=\prod_{j=1}^{d}  \Pr(Y_j=y_j)  \prod_{jk \in \mathcal{E}} \frac{  \Pr(Y_{j}=y_{j},Y_{k}=y_{k})}{\Pr(Y_j=y_j) \Pr(Y_k=y_k) }, 
\end{equation}
where $\Pr(Y_{j}=y_{j},Y_{k}=y_{k})= C_{jk}(a_{j,y_j+1},a_{k,y_k+1}) - C_{jk}(a_{j,y_j},  a_{k,y_k+1}) - C_{jk}(a_{j,y_j+1},a_{k,y_k}) + C_{jk}(a_{j,y_j},a_{k,y_k})$
and $\Pr(Y=y)=a_{j,y+1}-a_{j,y}.$

\subsection{Combined factor/truncated vine copula models}
In this section we combine the factor copula model with an 1-truncated vine copula to account for the residual dependence.
The pmf of an 1-truncated vine copula in (\ref{vine-pmf}) can be used in the pmf of the factor copula model in (\ref{p-factor-pmf}) instead of the product to capture any residual dependencies. Hence the pmf of the combined factor/truncated  vine copula model takes the form 
$$\pi_d(\y)=\int \prod_{j=1}^{d}  \Pr\left(Y_j=y_j|\X=\x\right)  \prod_{jk \in \mathcal{E}} \frac{  \Pr\left(Y_{j}=y_{j},Y_{k}=y_{k}|\X=\x\right)}{\Pr\left(Y_j=y_j|\X=\x\right) \Pr\left(Y_k=y_k|\X=\x\right) }\, dF_{\X}(\x).$$

With one factor and an 1-truncated vine given the latent variable $X_1$ (hereafter   1-factor tree)  let  $C_{jk;X_1}$ be a bivariate copula  such that
$$\Pr(Y_j\leq y_j,Y_k\leq y_k|X_1=x_1)=C_{jk;X_1}\bigl(F_{j|X_1}(y_j|x_1),F_{k|X_1}(y_k|x_1)\bigr),
$$
where $F_{j|X_1}$ and $F_{k|X_1}$ are given  in (\ref{eq-condjL1}). 
Then for a given 1-truncated vine structure with a set of edges $\mathcal{E}$, the pmf of the 1-factor tree copula model is
\begin{equation}\label{1-factor-vine-pmf}
\pi_d(\y)=\int_0^1
\prod_{j=1}^{d} f_{j|X_1}\left(y_j | x\right)  \prod_{jk \in \mathcal{E}}		 \frac{ f_{jk|X_1} (y_j,y_k|x_1)}{  f_{j|X}\left(y_j |x \right) f_{k|X}\left(y_k |x \right)}
 \, dx,
 \end{equation} 
where
$$
f_{jk|X_1}(y_j,y_k|x_1)=C_{jk;X_1} \bigl(  F_{j|X_1}^+,  F_{k|X_1}^+\bigr) - C_{jk;X_1} \bigl(  F_{j|X_1}^-,  F_{k|X_1}^+\bigr) \nonumber - C_{jk;X_1} \bigl(  F_{j|X_1}^+,  F_{k|X_1}^-\bigr) + C_{jk;X_1} \bigl(  F_{j|X_1}^-,  F_{k|X_1}^-\bigr)
$$
and $f_{j|X}\left(y_j |x \right)$, $f_{k|X}\left(y_k |x \right)$ are given in (\ref{1-factor-inner}). In the above  $F_{j|X_1}^+=F_{j|X_1}(y|x)$ and $F_{j|X_1}^-=F_{j|X_1}(y-1|x)$.

\begin{figure}[h!]
 \tiny
 \centering
\scalebox{0.8}{
    \begin{tikzpicture}
        [square/.style={
            draw,
            fill=gray!20,
            minimum width=3em,
            minimum height=3em,
            node contents={#1}}
            ]

        \node at (0,1) [circle,draw,fill=gray!20,minimum width=3em] (X) {\Large $X$};
           		       
	    \node at (-6,-2) (Y1) [square={\Large $Y_{1}$}]; 

        \node at (-3,-2) (Y2) [square={\Large $Y_{2}$}]; 

        \node at (0,-2) (Y3) [square={\Large $Y_{3}$}]; 
        
         \node at (3,-2) (Y4) [square={\Large $Y_{4}$}]; 
                
          \node at (6,-2) (Y5) [square={\Large $Y_{5}$}]; 

         	\node at  (-8,-2)  { \Large Tree 1 }; 
	    
\path(X)edge[black,bend right=0] node[above,rotate=30, midway,pos=0.55]{\large $Y_1X$}  (Y1);
\path(X)edge[black,bend right=0] node[above,rotate=45, midway,pos=0.55]{\large $Y_2X$} (Y2);
\path(X)edge[black,bend right=0] node[above,rotate=90, midway,pos=0.55]{\large $Y_3X$} (Y3);
\path(X)edge[black,bend right=0] node[above,rotate=-45, midway,pos=0.55]{\large $Y_4X$} (Y4);
\path(X)edge[black,bend right=0] node[above,rotate=-30, midway,pos=0.55]{\large $Y_5X$} (Y5);
	    \node at (-6,-4) (Y1) [square={\Large $Y_{1}X$}]; 

        \node at (-3,-4) (Y2) [square={\Large $Y_{2}X$}]; 

        \node at (0,-4) (Y3) [square={\Large $Y_{3}X$}]; 
        
         \node at (3,-4) (Y4) [square={\Large $Y_{4}X$}]; 
                
          \node at (6,-4) (Y5) [square={\Large $Y_{5}X$}]; 

         \node at  (-8,-4)  { \Large Tree 2 }; 
         	
\path(Y1)edge[black,bend right=35] node[above,rotate=0, midway,pos=0.55]{\large $Y_1Y_2|X$}  (Y2);
\path(Y2)edge[black,bend right=35] node[above,rotate=0, midway,pos=0.55]{\large $Y_2Y_3|X$} (Y3);
\path(Y2)edge[black,bend right=35] node[above,rotate=0, midway,pos=0.55]{\large $Y_2Y_4|X$} (Y4);
\path(Y4)edge[black,bend right=35] node[above,rotate=0, midway,pos=0.55]{\large $Y_4Y_5|X$} (Y5);

    \end{tikzpicture}
}
\caption{\label{fig:1FV} Graphical representation of a 1-factor tree copula model with $d=5$ items. The first tree is the 1-factor model. The residual dependence is captured in Tree 2 with an 1-truncated vine model.}
\end{figure}
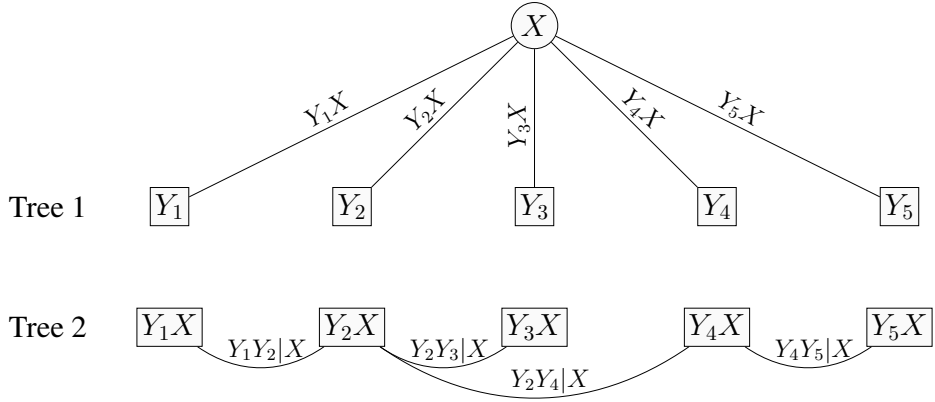

Figure \ref{fig:1FV}  depicts the graphical representation of a 1-factor tree copula model with $d=5$ items as a 2-truncated vine. Tree 1 shows the typical 1-factor model, while Tree 2 accounts for the residual dependence by   the pairwise conditional dependencies of  two items conditioned on the factor  $X_1$. 
 
With two factors and an 1-truncated vine given the latent variables $X_1,X_2$ (hereafter 2-factor tree),  let  $C_{jk;X_1,X_2}$ be a bivariate copula cdf such that 
$$\Pr(Y_j\leq y_j,Y_k\leq y_k|X_1,X_2)=C_{jk;X_1X_2}\bigl(F_{X_2j|X_1}(x_2,y_j|x_1),F_{X_2k|X_1}(x_2,y_k|x_1)\bigr),$$ where $F_{X_2j|X_1}$ and $F_{X_2k|X_1}$ are  given in (\ref{eq-condjL2}). 
Then for a given vine structure with a set of edges $\mathcal{E}$, the pmf of the 2-factor tree copula model is 
\begin{equation}\label{2-factor-vine-pmf}
\pi_d(\y)=\int_0^1\int_0^1
\prod_{j=1}^{d} f_{X_2j|X_1}\left(x_2,y_j | x_1\right)  \prod_{jk \in \mathcal{E}} \frac{f_{jk|X_1X_2}(y_j,y_k|x_1,x_2)}{ f_{X_2j|X_1}\left(x_2,y_j |x_1 \right) f_{X_2k|X_1}\left(x_2,y_k |x_1 \right) }
 \, d{x_1} d{x_2},
\end{equation}
 where 
\begin{align*} \label{eq:tildeC2}
f_{jk|X_1X_2}\bigl(y_j,y_k|x_1,x_2)=& C_{jk;X_1,X_2} \bigl(F_{X_2j|X_1}^+,  F_{X_2k|X_1}^+\bigr) - C_{jk;X_1,X_2} \bigl(  F_{X_2j|X_1}^-,  F_{X_2k|X_1}^+\bigr) \nonumber \\
&- C_{jk;X_1,X_2} \bigl(F_{X_2j|X_1}^+,F_{X_2k|X_1}^-\bigr) + C_{jk;X_1,X_2} \bigl(  F_{X_2j|X_1}^-,F_{X_2k|X_1}^-\bigr)
\end{align*}
and $f_{X_2j|X_1}(x_2,y_j|x_1)$, $f_{X_2k|X_1}(x_2,y_k|x_1)$ are as in (\ref{2-factor-inner}).  In the above $F_{X_2j|X_1}^+=F_{X_2j|X_1}(x_2,y|x_1)$ and $F_{X_2j|X_1}^-=F_{X_2j|X_1}(x_2,y-1|x_1)$. 
 
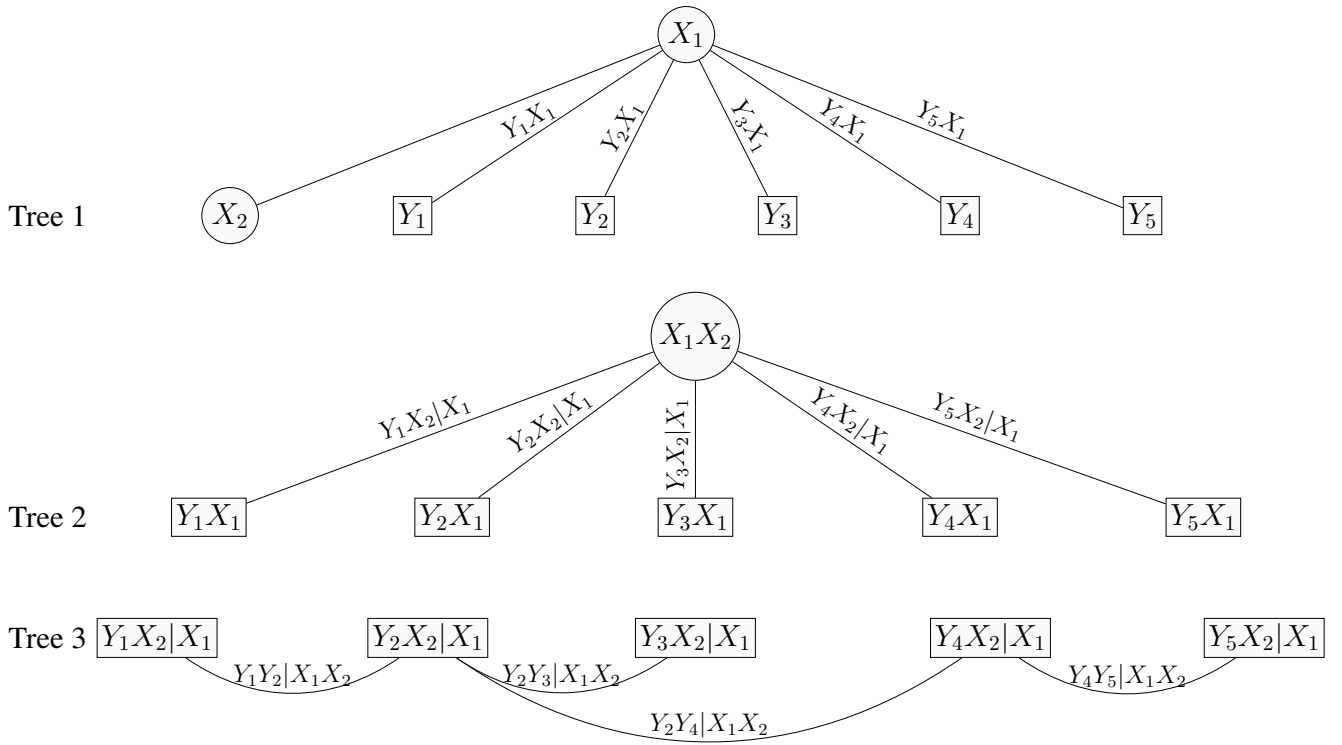
\begin{figure}[h!]
\tiny
\centering
\scalebox{0.8}{
    \begin{tikzpicture}
        [square/.style={
            draw,
            fill=gray!20,
            minimum width=3em,
            minimum height=3em,
            node contents={#1}}
            ]
            
        \node at (-1.5,1) [circle,draw,fill=gray!20,minimum width=3em] (X1) {\Large $X_1$};
           		       
	    \node at (-9,-2) [circle,draw,fill=gray!20,minimum width=3em] (X2) {\Large $X_2$}; 
	    
	    \node at (-6,-2) (Y1) [square={\Large $Y_{1}$}]; 

        \node at (-3,-2) (Y2) [square={\Large $Y_{2}$}]; 

        \node at (0,-2) (Y3) [square={\Large $Y_{3}$}]; 
        
         \node at (3,-2) (Y4) [square={\Large $Y_{4}$}]; 
                
          \node at (6,-2) (Y5) [square={\Large $Y_{5}$}]; 
	    
	      \node at  (-12,-2)  { \Large Tree 1 }; 
	             	
\path(X2)edge[black,bend right=0] node[above,rotate=30, midway,pos=0.55]{ }  (X1);

\path(X1)edge[black,bend right=0] node[above,rotate=30, midway,pos=0.55]{\large $Y_1X_1$}  (Y1);
\path(X1)edge[black,bend right=0] node[above,rotate=60, midway,pos=0.55]{\large $Y_2X_1$} (Y2);
\path(X1)edge[black,bend right=0] node[above,rotate=-60, midway,pos=0.55]{\large $Y_3X_1$} (Y3);
\path(X1)edge[black,bend right=0] node[above,rotate=-35, midway,pos=0.55]{\large $Y_4X_1$} (Y4);
\path(X1)edge[black,bend right=0] node[above,rotate=-25, midway,pos=0.55]{\large $Y_5X_1$} (Y5);

        \node at (-1.35,-4) [circle,draw,fill=gray!20,minimum width=3em] (X) {\Large $X_1X_2$};
           		       
	    \node at (-9.35,-7) (Y1) [square={\Large $Y_{1}X_1$}]; 

        \node at (-5.35,-7) (Y2) [square={\Large $Y_{2}X_1$}]; 

        \node at (-1.35,-7) (Y3) [square={\Large $Y_{3}X_1$}]; 
        
         \node at (3,-7) (Y4) [square={\Large $Y_{4}X_1$}]; 
                
          \node at (7,-7) (Y5) [square={\Large $Y_{5}X_1$}]; 

         	\node at  (-12,-7)  { \Large Tree 2 }; 
	    
\path(X)edge[black,bend right=0] node[above,rotate=20, midway,pos=0.55]{\large $Y_1X_2|X_1$}  (Y1);
\path(X)edge[black,bend right=0] node[above,rotate=35, midway,pos=0.55]{\large $Y_2X_2|X_1$} (Y2);
\path(X)edge[black,bend right=0] node[above,rotate=90, midway,pos=0.55]{\large $Y_3X_2|X_1$} (Y3);
\path(X)edge[black,bend right=0] node[above,rotate=-35, midway,pos=0.55]{\large $Y_4X_2|X_1$} (Y4);
\path(X)edge[black,bend right=0] node[above,rotate=-20, midway,pos=0.55]{\large $Y_5X_2|X_1$} (Y5);
	    \node at (-10.2,-9) (Y1) [square={\Large $Y_1X_2|X_1$}]; 

        \node at (-5.75,-9) (Y2) [square={\Large $Y_2X_2|X_1$}]; 

        \node at (-1.35,-9) (Y3) [square={\Large $Y_3X_2|X_1$}]; 
        
         \node at (3.5,-9) (Y4) [square={\Large $Y_4X_2|X_1$}]; 
                
          \node at (8,-9) (Y5) [square={\Large $Y_5X_2|X_1$}]; 

         \node at  (-12,-9)  { \Large Tree 3 }; 
         	
\path(Y1)edge[black,bend right=35] node[above,rotate=0, midway,pos=0.5]{\large $Y_1Y_2|X_1X_2$}  (Y2);
\path(Y2)edge[black,bend right=35] node[above,rotate=0, midway,pos=0.5]{\large $Y_2Y_3|X_1X_2$} (Y3);
\path(Y2)edge[black,bend right=35] node[above,rotate=0, midway,pos=0.5]{\large $Y_2Y_4|X_1X_2$} (Y4);
\path(Y4)edge[black,bend right=35] node[above,rotate=0, midway,pos=0.5]{\large $Y_4Y_5|X_1X_2$} (Y5);

    \end{tikzpicture}
}
\caption{\label{fig:2FV} Graphical representation of a 2-factor tree copula model with $d=5$ items. The first and second trees represent the 2-factor model. The residual dependence is captured in Tree 3 with an 1-truncated  vine model. Note that the factors are linked to one another with an independent copula in Tree 1.}
\end{figure}

Figure \ref{fig:2FV}  depicts the graphical representation of a 2-factor tree copula model with $d=5$ items as a 3-truncated vine. Trees 1 and 2 show the common 2-factor model, while  Tree 3 involves the pairwise conditional dependencies of two items given the factors.

For parametric 1-factor and 2-factor tree copula models, we let $C_{X_1j}$, 
$C_{X_2j}$ and $C_{jk;\X}$ be parametric bivariate copulas, say with parameters $\theta_{1j}$, $\theta_{2j}$, and $\delta_{jk}$, respectively. 
For the set of all parameters, let $\thbf=\{a_{jk}, \theta_{1j}, \delta_{jk}: j=1,\ldots,d;
k=1,\ldots,K-1;jk\in \mathcal{E} \}$
for the 1-factor tree copula model and
$\thbf=\{a_{jk}, \theta_{1j}, \theta_{2j}, \delta_{jk}: j=1,\ldots,d;
k=1,\ldots,K-1;jk\in \mathcal{E}\}$ for the
2-factor tree copula model.

\subsection{\label{other}Choices of parametric bivariate copulas}
In line with \cite{Nikoloulopoulos2015-PKA},  we use bivariate parametric copulas  that can be used when considering latent maxima, minima or mixtures of means. 
For different dependent items based on latent maxima or minima, multivariate extreme value and copula theory  (e.g., \citealt{Joe1997-CH} ) can be used to select suitable copulas that link observed to latent variables. 
Copulas that arise from extreme value theory have more probability in one joint tail (upper or lower) than expected with a discretized MVN distribution or a MVN copula with discrete margins. If item responses are based on discretizations of latent variables that are means, then it is possible that there can be more probability in both the joint upper and joint lower tail, compared with discretized MVN models. This happens if the respondents consist of a `mixture' population (e.g., different locations or genders). From the theory of elliptical distributions and copulas (e.g., \citealt{mcneil&frey&embrechts05}), it is known that the  multivariate Student-$t$ distribution as a  scale mixture of MVN has more dependence in the tails. Extreme value and elliptical  copulas can model item response data that have reflection asymmetric and symmetric dependence, respectively.

A bivariate copula $C$ is  reflection symmetric
if its density 
 satisfies $c(u_1,u_2)=c(1-u_1,1-u_2)$ for all $0\leq u_1,u_2\leq 1$.
Otherwise, it is reflection asymmetric often with more probability in the
joint upper tail or joint lower tail.  Upper tail dependence means
that $c(1-u,1-u)=O(u^{-1})$ as $u\to 0$ and  lower tail dependence
means that $c(u,u)=O(u^{-1})$ as $u\to 0$.
If $(U_1,U_2)\sim C$ for a bivariate copula $C$, then $(1-U_1,1-U_2)\sim
\widehat C$,  
where $\widehat C(u_1,u_2)=u_1+u_2-1+C(1-u_1,1-u_2)$   
is the survival or reflected 
copula of $C$; this ``reflection"
of each uniform $U(0,1)$ random variable about $1/2$ changes the direction
of tail asymmetry.  Choices of copulas with upper or lower tail dependence are better if the items have more probability in joint lower or upper tail than would be expected with the BVN copula.
This can be shown with summaries of polychoric correlations  in the upper and lower joint tail \citep{Kadhem&Nikoloulopoulos2021-BJMSP}.

After briefly providing definitions of  tail dependence  and reflection symmetry/asymmetry we provide below the bivariate  copula choices we consider: 

\begin{itemize}
\item The elliptical bivariate normal (BVN) copula with cdf
$$C(u_1,u_2;\th)=\Phi_2\Bigl(\Phi^{-1}(u_1;\nu),\Phi^{-1}(u_2;\nu);\th\Bigr),\hspace{2ex}-1\leq\th\leq 1,$$ 
where $\Phi$ is the univariate standard  normal cdf and  and $\Phi_2$ is the
cdf of a BVN  distribution with  correlation parameter $\th$.
A model with BVN  copulas   has latent (ordinal) variables that can be considered as  (discretized) means and and there is less  probability in both the joint upper and joint lower tail  as the BVN copula has reflection symmetry and  tail independence.
\item The extreme value Gumbel copula with cdf 
$$C(u_1,u_2;\th)=\exp\Bigl[-\Bigl\{(-\log u_1)^{\theta}
+(-\log u_2)^{\theta}\Bigr\}^{1/\theta}\Bigr],\hspace{2ex} \th\geq 1.$$
A  model with   bivariate  Gumbel copulas  has latent (ordinal) variables that can be considered as
(discretized) maxima and there is more probability in the joint upper tail  as the Gumbel copula has reflection asymmetry and  upper tail dependence.

\item The survival Gumbel (s.Gumbel) copula with cdf
$$C(u_1,u_2;\th)=u_1+u_2-1 + \exp\Bigl[-\Bigl\{\bigl(-\log (1-u_1)\bigr)^{\theta}
+\bigl(-\log (1-u_2)\bigr)^{\theta}\Bigr\}^{1/\theta}\Bigr],\hspace{2ex} \th\geq 1.$$ 
A  model with   bivariate  s.Gumbel copulas  has latent (ordinal) variables that can be considered as
(discretized) minima and there is more probability in the joint lower tail as the s.Gumbel copula has reflection asymmetry and  lower tail dependence.
\item The elliptical bivariate $t_\nu$ copula with cdf
$$C(u_1,u_2;\th)=\mathcal{T}_2\Bigl(\mathcal{T}^{-1}(u_1;\nu),\mathcal{T}^{-1}(u_2;\nu);\th,\nu\Bigr),\hspace{2ex}-1\leq\th\leq 1,$$ 
where $\mathcal{T}(;\nu)$ is the univariate Student-$t$ cdf with (non-integer) $\nu$ degrees of freedom, and $\mathcal{T}_2$ is the
cdf of a bivariate Student-$t$ distribution with $\nu$ degrees of freedom and correlation parameter $\th$.
A model with bivariate $t_\nu$ copulas   has latent (ordinal) variables that can be considered as mixtures of (discretized) means, since the bivariate Student-$t$ distribution arises as a scale mixture of bivariate normals. A small value of $\nu$, such as $1 \leq \nu\leq 5$, leads to a model with more probabilities in the joint upper and joint lower tails compared with the BVN copula   as the $t_\nu$ copula has reflection symmetric upper and  lower tail dependence.
\end{itemize}

For the residual part of the model  in addition to the aforementioned bivariate parametric copulas for computational improvements we can use the Archimedean  Frank copula with cdf 
$$C(u_1,u_2;\th)=-\theta^{-1}\log \left\{1+\frac{(e^{-\theta u_1}-1)(e^{-\theta
u_2}-1)}{e^{-\theta}-1} \right\}, \hspace{2ex} \th\in (-\infty,\infty)\setminus\{0\},$$
reflection symmetry and tail independence. 
Its  tail independence  is not a distributional concern about the tail dependence/asymmetry between the items  due to the main result in \citep{Joe&Li&Nikoloulopoulos2010}: for all the bivariate margins to have more probability in the joint lower or upper tail, it only suffices that the bivariate copulas in the first trees (factor part) to have upper/lower tail dependence and is not necessary for the bivariate copulas in the higher trees (residual part) to have tail dependence. For discrete data, such as item response,  the Frank copula has the same tail behaviour with the BVN copula but provides simplified computations  as it has a closed from cdf and thus it can preferred over the BVN  copula for the residual part of the model that involves finite  differences of bivariate copula cdfs.

\section{Estimation}\label{sec:Estimation}

With sample size $n$ and data $\y_1,\ldots,\y_n$, the  joint log-likelihood of the  factor tree copula models is 
\begin{equation}\label{joint-loglik}
\ell(\thbf;\y_1,\ldots,\y_n)=\sum_{i=1}^n\log \pi_d (\y_i;\thbf),
\end{equation}
with $\pi_d(\y)$ as defined in (\ref{1-factor-vine-pmf}) and (\ref{2-factor-vine-pmf}) for the 1-factor and 2-factor tree copula model, respectively. 
Maximization of (\ref{joint-loglik})
is numerically possible
but time-consuming  for large $d$ because of many univariate cutpoints and
dependence parameters.
Hence, we approach estimation using the  two-step IFM method proposed  by \cite{Joe2005-JMVA} that 
can efficiently, in the sense of computing time and asymptotic variance,
estimate the model parameters.

In the first step, the cutpoints are estimated using the univariate sample proportions. The univariate cutpoints for the $j$th item are estimated as $\hat{a}_{j,k} = \sum_{y=0}^{k} p_{j,y}$, where  $p_{j,y}\,,y=0,\ldots,K-1$ for $j=1,\ldots,d$ are the univariate sample proportions. 
In the second step of the IFM method, the joint log-likelihood in (\ref{joint-loglik}) is maximized over the copula parameters with the cutpoints fixed as estimated at the first step. The estimated copula parameters can be obtained by using a quasi-Newton \citep{Nash1990} method applied to the logarithm of the joint likelihood.

For the 1-factor tree copula model, numerical evaluation of the joint pmf can be achieved with the following steps:

\begin{enumerate}
\itemsep=10pt
\item Calculate Gauss-Legendre quadrature \citep{Stroud&Secrest1966} points  $\{x_q: q=1,\ldots,n_q\}$  and weights $\{w_q: q=1,\ldots,n_q\}$  in terms of standard uniform.  

\item Numerically evaluate the joint pmf in (\ref{1-factor-vine-pmf}) via the following approximation:
$$
\sum_{q=1}^{n_q} w_{q}
\prod_{j=1}^{d} f_{j}(y_j | x_q)  \prod_{[jk] \in \mathcal{E}} 
\frac{f_{jk|X_1} (y_j,y_k|x_q)}{  f_{j|X}(y_j |x_q ) f_{k|X}(y_k |x_q)}.$$
\end{enumerate}

For the 2-factor tree copula model, numerical evaluation of the joint pmf can be achieved with the following steps:

\begin{enumerate}
\itemsep=10pt
\item Calculate Gauss-Legendre quadrature \citep{Stroud&Secrest1966} points  $\{x_{q_1}: q_1 = 1,\ldots,n_q\}$ and  $\{x_{q_2}: q_2 = 1,\ldots,n_q\}$   and weights $\{w_{q_1}: q_1 = 1,\ldots,n_q\}$ and  $\{w_{q_2}: q_2 = 1,\ldots,n_q\}$   in terms of standard uniform.  

\item Numerically evaluate the joint pmf in (\ref{2-factor-vine-pmf}) via the following approximation in a double sum: 
$$\sum_{q_1=1}^{n_q} \sum_{q_2=1}^{n_q}  w_{q_1} w_{q_2} 
 \prod_{j=1}^{d} f_{X_2j|X_1}(x_{q_2} ,y_j | x_{q_1})  
 \prod_{[jk] \in \mathcal{E}} \frac{f_{jk|X_1X_2}(y_j,y_k|x_{q_1},x_{q_2})}{f_{X_2j|X_1}(x_{q_2} ,y_j|x_{q_1}) f_{X_2k|X_1}(x_{q_2} ,y_k |x_{q_1} )}.$$

\end{enumerate}

With Gauss-Legendre quadrature, the same nodes and weights are used for different functions; this helps in
yielding smooth numerical derivatives for numerical optimization via quasi-Newton. Our comparisons show that $n_q=15$ quadrature points are adequate with good precision.

\section{Model selection}\label{sec:ModelSelection}
In this section we will discuss  model selection strategies for the factor tree copula models. Section \ref{sec:vinetree} proposes   vine tree structure selection methods for the residual part of the model that assume  the factor tree copula  models are constructed with bivariate normal (BVN) copulas. Section \ref{sec:SelectingCopulas}  proposes a heuristic algorithm that  sequentially selects suitable bivariate copulas to account for any tail dependence/asymmetry as in \cite{Kadhem&Nikoloulopoulos2021,Kadhem&Nikoloulopoulos2021-BJMSP}. 

\subsection{1-truncated vine tree structure selection } \label{sec:vinetree}
We propose two selection algorithms to choose the 1-truncated vine tree structure  $ \mathcal{E}$ for the residual part of the model, namely  the polychoric and partial selection algorithms.  Before that, we provide the necessary  tools to form the aforementioned algorithms. These are the estimated  polychoric correlations \citep{Olsson-1979}, correlations between each of the items and the first factor and partial correlations  between each of the items and the second  factor  given the first factor \citep{Nikoloulopoulos2015-PKA}.

The  sample polychoric correlation for all possible pairs  of items can be estimated  as 
\begin{align*}\label{polychoric}
\hat\rho_{jk}=\mbox{argmax}_\rho\sum_{i=1}^n&\log\Bigl(\Phi_2(\alpha_{j,y_{ij}+1},\alpha_{k,y_{ik}+1}; \rho)-\Phi_2(\alpha_{j,y_{ij}+1},\alpha_{k,y_{ik}}; \rho)-\nonumber\\&\Phi_2(\alpha_{j,y_{ij}},\alpha_{k,y_{ik}+1}; \rho)+\Phi_2(\alpha_{j,y_{ij}},\alpha_{k,y_{ik}}; \rho)\Bigr),\quad 1\leq j<k\leq d,
\end{align*}
where $\Phi_2(\cdot,\cdot;\rho)$ is the BVN cdf with correlation parameter $\rho$.

When all the bivariate copulas are BVN  the $p$-factor copula model is the same as the discretized MVN model with a $p$-factor correlation matrix, also known as the $p$-dimensional normal ogive model \citep{Joreskog&Moustaki2001}. 
The 1-factor copula model in (\ref{1-pdf}) is the same as the variant of Samejima's (1969) \nocite{Samejima-1969} graded response IRT model, known as normal ogive model \citep{McDonald-1997} with a 1-factor correlation
matrix $R=(r_{jk})$ with $r_{jk}=\th_{1j}\th_{1k}$ for $j\ne k$.
The 2-factor model in (\ref{2-pdf})  is the same as the bidimensional (2-factor) normal ogive model with a 2-factor correlation
matrix $R=(r_{jk})$ with $r_{jk}=\th_{1j}\th_{1k}+
\th_{2j}\th_{2k}[(1-\th_{1j}^2)(1-\th_{1k}^2)]^{1/2}$ for $j\ne k$.
The parameter $\theta_{1j}$ of $C_{X_1j}$ is the correlation of  the underlying normal variable $Z_j$ of $Y_j$ with $Z_{01}=\Phi^{-1}(X_1)$, and
the parameter $\th_{2j}$ of $C_{X_2j}$ is the partial correlation
between $Z_j$   and $Z_{02}=\Phi^{-1}(X_1)$ given $Z_{01}$.

Subsequently, for all possible pair of items we can estimate the 
partial correlations between $Z_j$ and $Z_k$ given $Z_{01}$  and the partial correlations  between $Z_j$ and $Z_k$  given $Z_{01},Z_{02}$ 
via the relations 
$$
\hat \rho_{jk;Z_{01}}  =  \frac{\hat\rho_{jk}  -\hat \theta_{1j} \hat\theta_{1k} }{  \sqrt{ (1-\hat\theta_{1j}^2) (1-\hat\theta_{1k}^2) } }
\quad \mbox{and}\quad 
\hat\rho_{jk;Z_{01},Z_{02}} = \frac{  \hat\rho_{jk;Z_{01}}  - \hat\th_{2j} \hat\th_{2k} }{  \sqrt{ (1-\hat\th_{2j}^2) (1-\hat\th_{2k}^2) } },
$$
respectively, where  $\hat\theta_{1j},\hat\theta_{1k}$ are the estimated unidimensional normal ogive model's parameters and $\hat\theta_{1j},\hat\theta_{1k},\hat\theta_{2j},\hat\theta_{2k}$  are the estimated bidimensional normal ogive model's parameters.

The polychoric and partial correlation algorithms  select the best vine tree using the minimum spanning tree algorithm \citep{Prim1957}. The former algorithm selects the edges  $\mathcal{E}$ of the tree that   minimize the sum of the 
weights $\log(1-\hat\rho_{jk}^2)$, while the latter  algorithm the  sum of the weights $\log(1-\hat\rho_{jk;Z_{01}}^2)$ for the 1-factor tree copula model    and $\log(1-\hat\rho_{jk;Z_{01},Z_{02}}^2)$ for the 2-factor tree copula model.

\subsection{Bivariate copula selection} \label{sec:SelectingCopulas}
We propose a heuristic method that  selects appropriate bivariate copulas for  the proposed  models. It starts with an initial assumption that all bivariate copulas are BVN and independent copulas in the factor  and 1-truncated vine copula model, respectively. Then sequentially suitable copulas with lower or upper tail dependence are assigned 
where necessary to account for more probability in one or both joint tails. 
For ease of interpretation, we do not mix Gumbel, s.Gumbel, $t_\nu$ and BVN for a single tree of the model; e.g., for the 2-factor tree copula model we allow three different copula families, one for the first factor, one for the second factor and one for the 1-truncated vine (residual dependence  part of the model).

The selection algorithm involves the following steps:

\begin{enumerate}

\item Start with a factor tree copula model with   BVN and independent copulas in the factor  and 1-truncated vine copula parts of the model, respectively. 

\item Factor part
\begin{enumerate}

\item Factor 1
\begin{enumerate}
\item Fit all the possible  models, iterating over all the bivariate copula candidates that link each of the items to $X_1$. 
\item Select the bivariate copula  that corresponds to the highest log-likelihood. 

\item Replace the BVN with the selected bivariate copula that links each of  the items to $X_1$.
\end{enumerate}

\item Factor 2
\begin{enumerate}
\item  Fit all the possible models, iterating over all the copula candidates that link each of the items  to $X_2$. 
\item Select the  bivariate copula that corresponds to the highest log-likelihood. 
\item Replace BVN with  the selected bivariate copula that links each of the items  to  $X_2$.
\end{enumerate}
\end{enumerate}
\item 1-truncated vine part
\begin{enumerate}

\item Select the best 1-truncated vine tree structure $\mathcal{E}$  using  both the polychoric and partial selection algorithms proposed in  Subsection \ref{sec:vinetree}.

\item Fit all the possible  models, iterating over all the bivariate copula candidates that link the pairs of  items $\in \mathcal{E}$  given the factors. 

\item Select  the bivariate  copula  that corresponds to the highest log-likelihood.

\item Replace the independence copula with the selected bivariate copula that links each pair of items $\in \mathcal{E}$ given the factors.

\end{enumerate}
\end{enumerate}

\begin{sidewaystable}[htbp]
  \centering
  \footnotesize
  \caption{Small sample of size $n = 500$ simulations ($10^3$ replications) and $d=\{8,16,24\}$ items with $K=5$ equally weighted categories from an 1-factor tree  copula model with Gumbel copulas and  an 1-truncated  drawable vine residual dependence structure for $d=\{8, 16, 24\}$ and resultant  biases, root mean square errors (RMSE), and standard deviations (SD), scaled by $n$, for the IFM estimates. }

    \setlength{\tabcolsep}{10.6pt}  
            
    \begin{tabular}{lcccccccccccccccc}
    \toprule
 $d=8$     & \multicolumn{8}{c}{1st tree (1-factor copula)} &   & \multicolumn{7}{c}{2nd tree  (1-truncated drawable vine copula)} \\
\cmidrule{2-9}\cmidrule{11-17}    $\tau$ & 0.70 & 0.66 & 0.61 & 0.57 & 0.53 & 0.49 & 0.44 & 0.40 &   & 0.40 & 0.35 & 0.30 & 0.25 & 0.20 & 0.15 & 0.10 \\
    \midrule
    $n$Bias & 6.19 & 5.83 & 8.34 & 7.30 & 4.13 & -0.46 & -2.47 & -2.77 &   & -14.23 & -16.11 & -15.79 & -9.90 & -2.86 & 1.19 & 1.42 \\
    $n$SD & 20.48 & 21.24 & 19.05 & 17.56 & 16.43 & 16.56 & 15.79 & 16.05 &   & 44.97 & 33.61 & 28.66 & 25.17 & 21.68 & 19.87 & 18.54 \\
    $n$RMSE & 21.40 & 22.03 & 20.80 & 19.01 & 16.94 & 16.57 & 15.98 & 16.29 &   & 47.17 & 37.27 & 32.72 & 27.05 & 21.87 & 19.91 & 18.60 \\
    \midrule
$d=16$      & \multicolumn{16}{c}{1st tree (1-factor copula)} \\
\cmidrule{2-17}    $\tau$ & 0.70 & 0.68 & 0.66 & 0.64 & 0.62 & 0.60 & 0.58 & 0.56 & 0.54 & 0.52 & 0.50 & 0.48 & 0.46 & 0.44 & 0.42 & 0.40 \\
    \midrule
    $n$Bias & 2.76 & 3.43 & 5.22 & 6.18 & 6.02 & 4.66 & 2.96 & 2.19 & 0.79 & 0.20 & 0.05 & -1.43 & -1.74 & -1.02 & -1.80 & -0.93 \\
    $n$SD & 10.89 & 11.31 & 11.85 & 11.94 & 12.08 & 11.91 & 12.35 & 12.45 & 12.65 & 13.26 & 12.96 & 13.66 & 13.66 & 14.51 & 14.55 & 14.19 \\
    $n$RMSE & 11.23 & 11.81 & 12.95 & 13.45 & 13.49 & 12.79 & 12.70 & 12.64 & 12.68 & 13.26 & 12.96 & 13.74 & 13.77 & 14.55 & 14.66 & 14.22 \\
    \midrule
      & \multicolumn{16}{c}{2nd tree (1-truncated drawable vine copula)} \\
\cmidrule{2-17}    $\tau$ & 0.40 & 0.38 & 0.36 & 0.34 & 0.31 & 0.29 & 0.27 & 0.25 & 0.23 & 0.21 & 0.19 & 0.16 & 0.14 & 0.12 & 0.10 &  \\
    \midrule
    $n$Bias & -6.55 & -9.58 & -12.27 & -11.32 & -9.85 & -6.42 & -4.51 & -2.46 & -1.01 & 0.46 & 0.70 & 1.35 & 1.96 & 1.17 & 1.59 &  \\
    $n$SD & 22.62 & 22.71 & 21.92 & 20.66 & 19.36 & 18.59 & 18.95 & 18.22 & 17.92 & 18.02 & 17.21 & 17.20 & 16.79 & 16.91 & 16.62 &  \\
    $n$RMSE & 23.55 & 24.65 & 25.12 & 23.56 & 21.72 & 19.67 & 19.48 & 18.39 & 17.95 & 18.02 & 17.22 & 17.25 & 16.90 & 16.95 & 16.70 &  \\
    \bottomrule
    \end{tabular}

    \setlength{\tabcolsep}{4.3pt}  
    
    \begin{tabular}{lcccccccccccccccccccccccc}
    
$d=24$      & \multicolumn{24}{c}{1st tree (1-factor copula)} \\
\cmidrule{2-25}    $\tau$ & 0.70 & 0.69 & 0.67 & 0.66 & 0.65 & 0.63 & 0.62 & 0.61 & 0.60 & 0.58 & 0.57 & 0.56 & 0.54 & 0.53 & 0.52 & 0.50 & 0.49 & 0.48 & 0.47 & 0.45 & 0.44 & 0.43 & 0.41 & 0.40 \\
    \midrule
    $n$Bias & 1.61 & 1.89 & 3.41 & 4.20 & 4.35 & 3.84 & 3.13 & 2.52 & 2.29 & 1.68 & 1.03 & 0.44 & -0.21 & 0.05 & -0.53 & -0.55 & -0.28 & -0.05 & -0.12 & -0.33 & -0.44 & -0.12 & -0.25 & -0.60 \\
    $n$SD & 9.72 & 10.39 & 10.86 & 11.06 & 11.13 & 10.86 & 11.28 & 11.32 & 11.61 & 11.99 & 11.76 & 11.90 & 12.10 & 12.54 & 12.71 & 12.70 & 12.82 & 13.21 & 13.54 & 13.43 & 13.86 & 13.74 & 13.57 & 13.84 \\
    $n$RMSE & 9.86 & 10.56 & 11.38 & 11.83 & 11.95 & 11.52 & 11.70 & 11.59 & 11.83 & 12.11 & 11.80 & 11.91 & 12.11 & 12.54 & 12.73 & 12.71 & 12.82 & 13.21 & 13.54 & 13.43 & 13.87 & 13.74 & 13.58 & 13.85 \\
        \midrule
      & \multicolumn{24}{c}{2nd tree  (1-truncated drawable vine copula)} \\
\cmidrule{2-25}    $\tau$ & 0.40 & 0.39 & 0.37 & 0.36 & 0.35 & 0.33 & 0.32 & 0.30 & 0.29 & 0.28 & 0.26 & 0.25 & 0.24 & 0.22 & 0.21 & 0.20 & 0.18 & 0.17 & 0.15 & 0.14 & 0.13 & 0.11 & 0.10 &  \\
    \midrule
    $n$Bias & -4.29 & -6.22 & -7.94 & -8.53 & -7.72 & -6.13 & -6.24 & -4.19 & -2.61 & -2.03 & -1.33 & -0.34 & 0.17 & -0.32 & 0.59 & -0.06 & 0.44 & 1.32 & 0.74 & 0.60 & 0.46 & 0.04 & 0.73 &  \\
    $n$SD & 20.39 & 19.93 & 19.75 & 19.11 & 19.40 & 18.36 & 18.93 & 18.26 & 18.15 & 18.02 & 17.04 & 17.40 & 16.77 & 17.03 & 17.64 & 16.52 & 17.28 & 16.69 & 17.22 & 16.72 & 17.12 & 16.88 & 16.79 &  \\
    $n$RMSE & 20.84 & 20.88 & 21.28 & 20.93 & 20.88 & 19.36 & 19.94 & 18.73 & 18.33 & 18.14 & 17.10 & 17.41 & 16.78 & 17.04 & 17.65 & 16.52 & 17.29 & 16.74 & 17.24 & 16.73 & 17.13 & 16.88 & 16.80 &  \\
    \bottomrule
    \end{tabular}
\label{tab:sim.1fvine}
\end{sidewaystable}

\section{Simulations}\label{sec:Simulations}
An extensive simulation study is conducted to assess the (a) efficiency of the proposed estimation method  and (b) reliability of using the model selection algorithms to select the correct 1-truncated vine tree structure for the residual dependence part of the model. 
We randomly generated $1,000$ datasets with sample size $n = 500$ and $d=\{8,16,24\}$ items with $K=5$ equally weighted categories from an 1-factor and 2-factor tree copula models with  Gumbel  copulas. The items   in the last tree  are either serially connected in ascending order with an 1-truncated drawable vine  or randomly connected with a 1-truncated regular vine. Note in passing that the drawable vine is a boundary  regular vine case.

We set the copula parameters in Kendall's $\tau$ scale, i.e., $\tau(\th_{1j},\,j=1,\ldots,d)=\{0.70,\ldots,0.40 \}$ and  $\tau(\th_{2j},\,\,j=1,\ldots,d)=\{0.55,\ldots,0.25 \}$ for the factor copula parts of the models and  $\tau(\delta_{jk},\,jk\in \mathcal{E})=\{0.55,\ldots,0.25 \}$ and $\tau(\delta_{jk},\,jk\in \mathcal{E})=\{0.40,\ldots,0.10 \}$ for the  1-truncated vine copula part of the model for the 1-factor and 2-factor tree copula model, respectively. The $\tau$'s as above form equally spaced sequences and are strictly increasing functions of the  
true (simulated) Gumbel 
 copula parameters, viz.  
\begin{equation}\label{tauGumbel}
\tau(\th)=1-\th^{-1}.
\end{equation}

Table \ref{tab:sim.1fvine} and Table \ref{tab:sim.2fvine}  present the resulting biases, standard deviations (SD) and root mean square errors (RMSE), scaled by $n$, from the simulations of  the 1-factor and  2-factor tree copula models with Gumbel copulas, respectively and  an 1-truncated  drawable vine residual dependence structure.  The results indicate that the proposed approximation method is efficient for estimating the factor tree  copula models  and the efficiency  improves as the dimension increases.

\begin{sidewaystable}[htbp]
    \footnotesize
        \setlength{\tabcolsep}{3.65pt}  
                       \renewcommand{\arraystretch}{1.2}		
  \caption{Small sample of size $n = 500$ simulations ($10^3$ replications) and $d=24$ items with $K=5$ equally weighted categories from a 2-factor tree  copula model with Gumbel copulas and  an 1-truncated  drawable vine residual dependence structure  and resultant  biases, root mean square errors (RMSE), and standard deviations (SD), scaled by $n$, for the IFM estimates.}
    \begin{tabular}{lcccccccccccccccccccccccc}
    \toprule
    \multicolumn{25}{l}{$d=24$} \\
    \midrule
    \multicolumn{1}{c}{ } & \multicolumn{24}{c}{1st tree (1st factor of 2-factor copula)} \\
\cmidrule{2-25}    $\tau$ & 0.70 & 0.69 & 0.67 & 0.66 & 0.65 & 0.63 & 0.62 & 0.61 & 0.60 & 0.58 & 0.57 & 0.56 & 0.54 & 0.53 & 0.52 & 0.50 & 0.49 & 0.48 & 0.47 & 0.45 & 0.44 & 0.43 & 0.41 & 0.40 \\
    \midrule
    $n$Bias & -5.74 & -3.26 & -0.07 & 2.35 & 3.96 & 4.12 & 3.60 & 3.94 & 4.05 & 3.73 & 4.58 & 4.27 & 3.74 & 4.83 & 4.17 & 5.08 & 4.28 & 4.56 & 5.15 & 4.80 & 4.82 & 4.05 & 4.42 & 2.96 \\
    $n$SD & 26.55 & 26.96 & 27.90 & 27.43 & 25.80 & 24.89 & 25.15 & 24.57 & 23.62 & 23.93 & 23.89 & 23.53 & 23.21 & 23.04 & 22.38 & 23.15 & 22.39 & 23.75 & 22.93 & 22.04 & 22.38 & 21.99 & 22.71 & 21.74 \\
    $n$RMSE & 27.16 & 27.15 & 27.90 & 27.53 & 26.11 & 25.23 & 25.41 & 24.89 & 23.97 & 24.22 & 24.33 & 23.91 & 23.51 & 23.54 & 22.77 & 23.70 & 22.80 & 24.18 & 23.50 & 22.56 & 22.89 & 22.36 & 23.14 & 21.94 \\
    \midrule
    \multicolumn{1}{c}{ } & \multicolumn{24}{c}{2nd tree (2nd factor of 2-factor copula)} \\
\cmidrule{2-25}    $\tau$ & 0.55 & 0.54 & 0.52 & 0.51 & 0.50 & 0.48 & 0.47 & 0.46 & 0.45 & 0.43 & 0.42 & 0.41 & 0.39 & 0.38 & 0.37 & 0.35 & 0.34 & 0.33 & 0.32 & 0.30 & 0.29 & 0.28 & 0.26 & 0.25 \\
    \midrule
    $n$Bias & 4.31 & 1.24 & 2.81 & 0.39 & -0.58 & -1.81 & -2.58 & -3.06 & -6.03 & -6.58 & -8.23 & -9.13 & -9.58 & -12.73 & -13.14 & -11.90 & -9.67 & -10.48 & -12.89 & -11.57 & -11.57 & -12.77 & -11.14 & -8.04 \\
    $n$SD & 40.65 & 41.80 & 42.93 & 45.05 & 43.16 & 42.69 & 41.67 & 40.68 & 40.38 & 41.00 & 41.35 & 39.73 & 41.24 & 41.35 & 40.48 & 40.60 & 41.84 & 42.41 & 40.90 & 38.62 & 40.15 & 37.78 & 39.96 & 38.41 \\
    $n$RMSE & 40.88 & 41.82 & 43.02 & 45.05 & 43.17 & 42.73 & 41.75 & 40.79 & 40.83 & 41.52 & 42.16 & 40.76 & 42.34 & 43.27 & 42.56 & 42.31 & 42.94 & 43.68 & 42.88 & 40.31 & 41.78 & 39.88 & 41.49 & 39.25 \\
    \midrule
      & \multicolumn{24}{c}{3rd tree  (1-truncated drawable vine copula)} \\
\cmidrule{2-25}    $\tau$ & 0.40 & 0.39 & 0.37 & 0.36 & 0.35 & 0.33 & 0.32 & 0.30 & 0.29 & 0.28 & 0.26 & 0.25 & 0.24 & 0.22 & 0.21 & 0.20 & 0.18 & 0.17 & 0.15 & 0.14 & 0.13 & 0.11 & 0.10 &  \\
    \midrule
    $n$Bias & 0.10 & -4.49 & -9.56 & -10.74 & -9.52 & -9.21 & -6.47 & -4.90 & -2.94 & -3.25 & -0.50 & -0.21 & 0.85 & 1.52 & 2.04 & 0.34 & 1.66 & 1.66 & 1.76 & 2.45 & 2.02 & 2.29 & 2.25 &  \\
    $n$SD & 32.64 & 35.17 & 31.46 & 28.61 & 27.74 & 24.35 & 24.49 & 22.53 & 25.08 & 23.54 & 22.79 & 20.38 & 21.06 & 20.56 & 20.37 & 22.01 & 20.16 & 20.08 & 19.14 & 19.56 & 18.21 & 18.11 & 18.33 &  \\
    $n$RMSE & 32.64 & 35.46 & 32.88 & 30.56 & 29.33 & 26.03 & 25.33 & 23.06 & 25.25 & 23.76 & 22.80 & 20.38 & 21.07 & 20.61 & 20.48 & 22.01 & 20.23 & 20.15 & 19.22 & 19.71 & 18.33 & 18.25 & 18.47 &  \\
    \bottomrule
    \end{tabular}
\label{tab:sim.2fvine}
\end{sidewaystable}

\begin{landscape}
\begin{figure}[!h]
\caption{\label{sel-times}Small sample of size $n = 500$ simulations ($10^3$ replications) and $d=\{8,16,24\}$ items with $K=5$ equally weighted categories from 1-factor  and 2-factor tree copula models with Gumbel copulas and  an 1-truncated  drawable/regular  vine residual dependence structure and resultant 
number of times a pair of items is correctly selected as an edge for each of the edges of the 1-truncated drawable and regular vine  copula for both the partial and polychoric selection algorithms. 
 }
\begin{center}
\begin{tabular}{ccc}
\hline
\multicolumn{3}{c}{1-factor tree copula models}\\
$d=8$ &$d=16$ & $d=24$\\\hline
\includegraphics[width=0.5\textwidth]{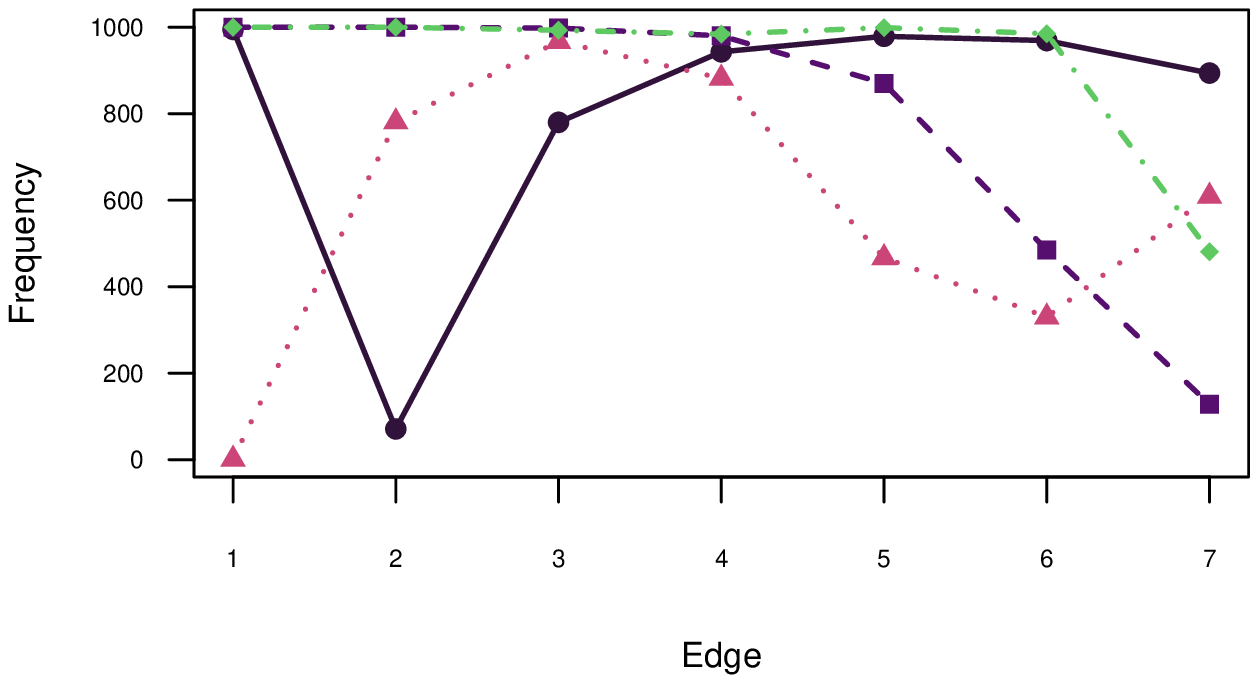} &
\includegraphics[width=0.5\textwidth]{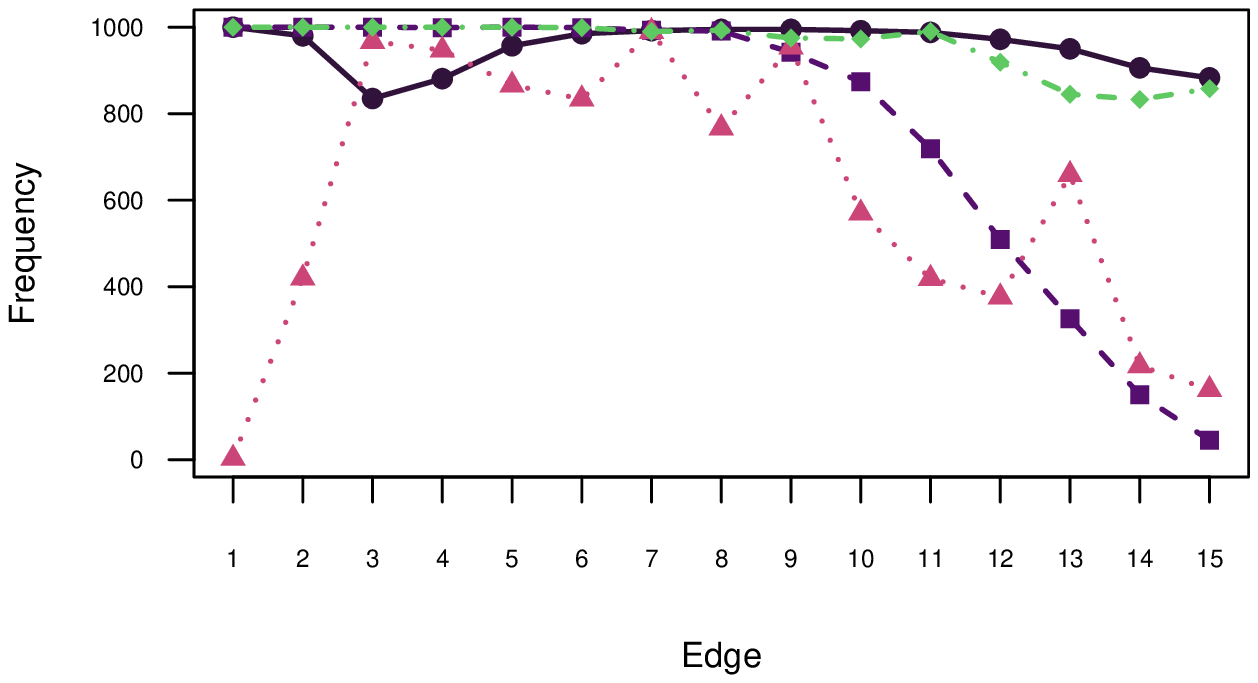}&
\includegraphics[width=0.5\textwidth]{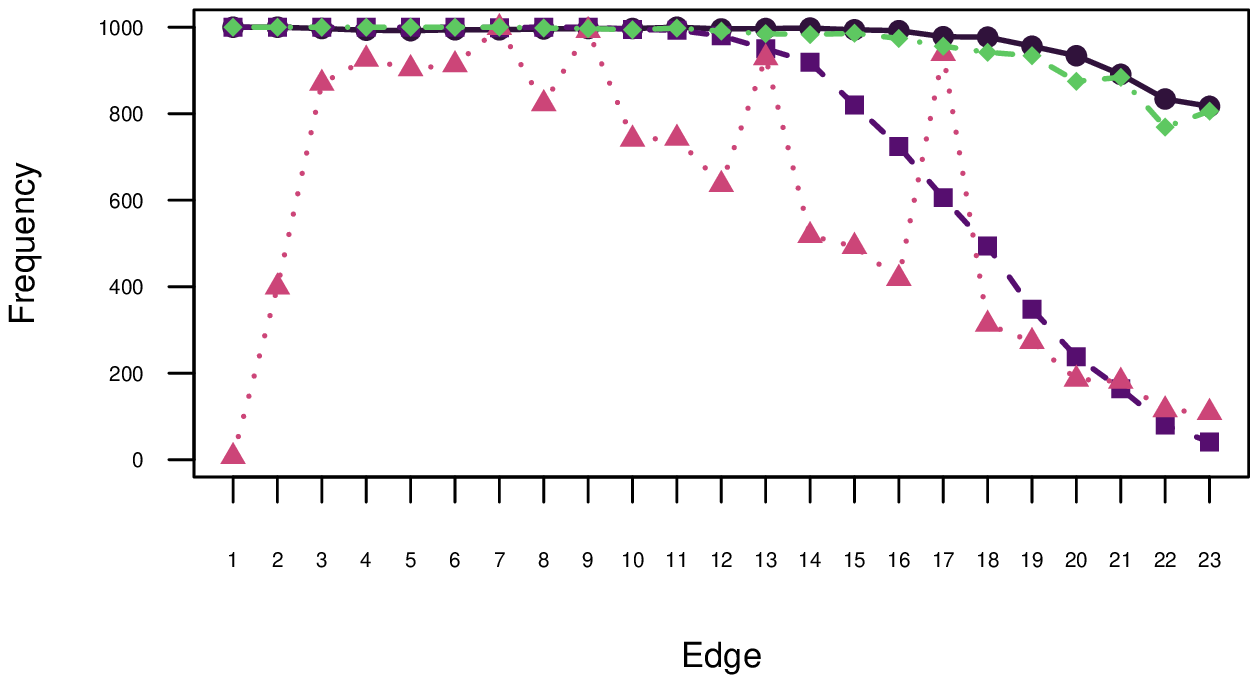}\\
\hline
\multicolumn{3}{c}{2-factor tree copula models}\\
$d=8$ &$d=16$ & $d=24$\\\hline
\includegraphics[width=0.5\textwidth]{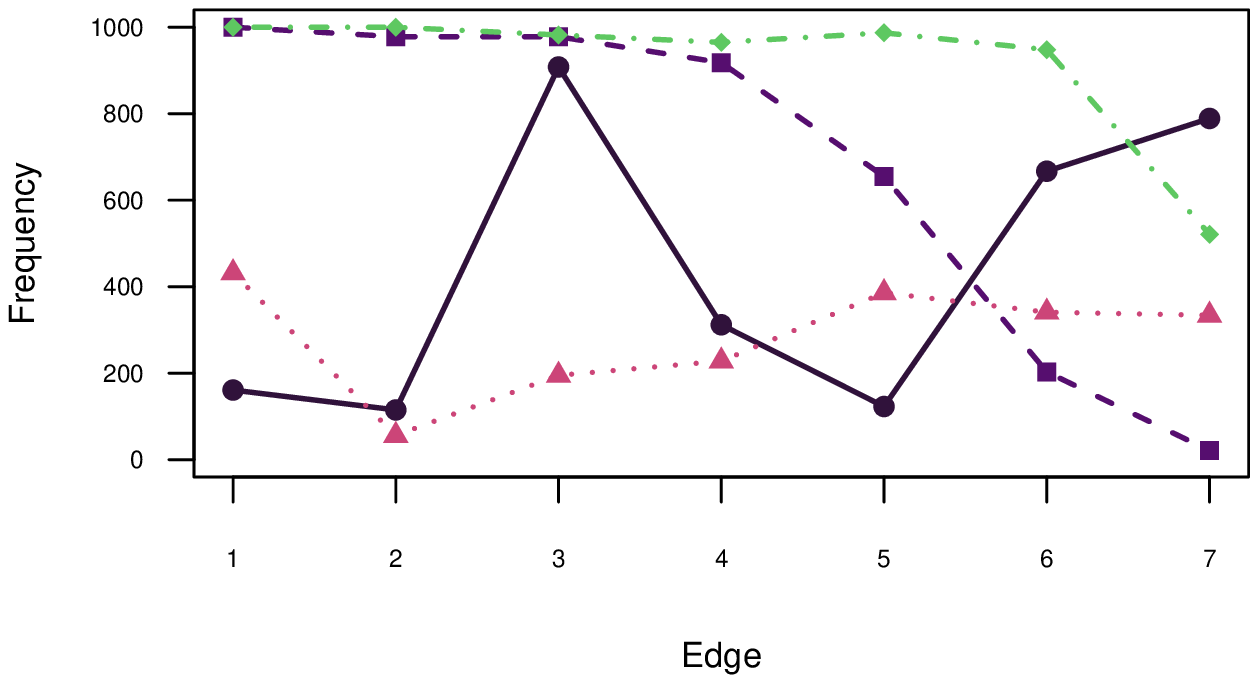} &
\includegraphics[width=0.5\textwidth]{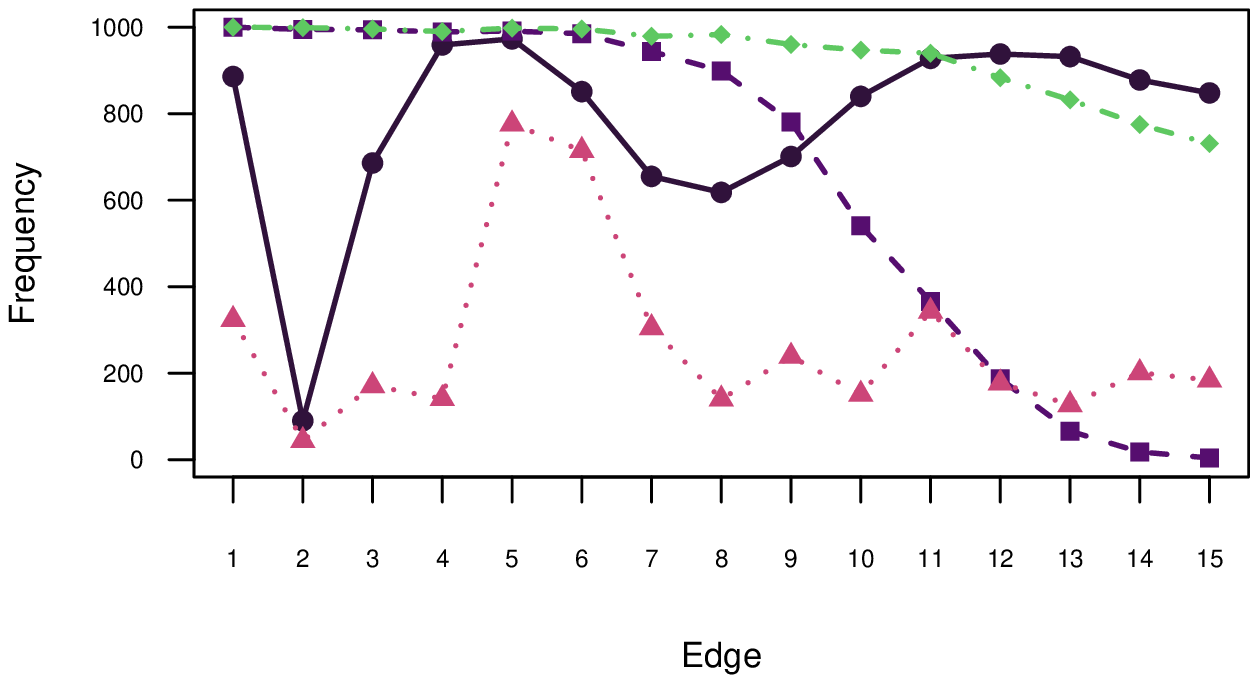}&
\includegraphics[width=0.5\textwidth]{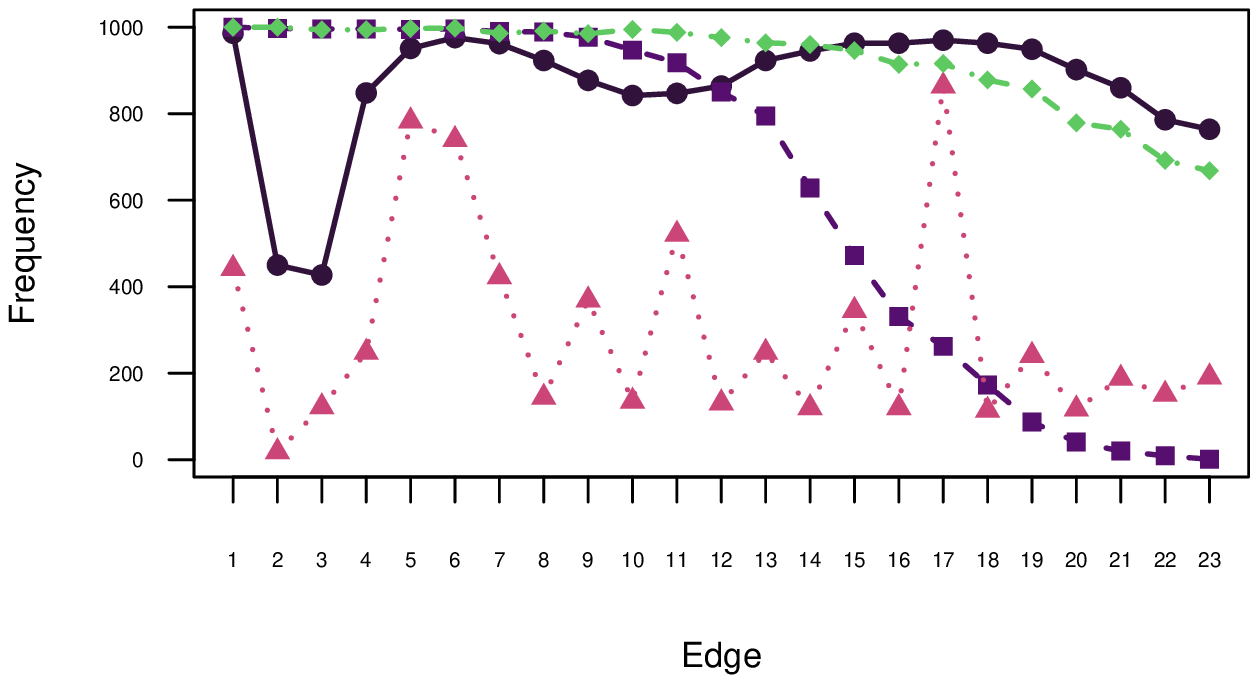}\\\hline
\multicolumn{3}{c}{Drawable vine partial: (\protect\blackcircle)\hspace{10ex} Drawable vine polychoric: (\protect\violetrectangle)\hspace{10ex}   Regular vine partial: (\protect\pinktriangle)\hspace{10ex} Regular vine polychoric: (\protect\greendiamond)}\\
\hline
\end{tabular}
\end{center}

\end{figure}
\end{landscape}

In Figure \ref{sel-times} we report the frequency of a pair of items is correctly selected as an edge for each of the edges of the 1-truncated vine from the simulations of the 1- and 2-factor tree copula models with Gumbel copulas with $d=8$, $d=16$ and $d=24$ items  for both the partial and polychoric selection algorithms. It has been shown  that the partial selection algorithm as the dimension  increases  performs extremely well for  the 1-truncated  drawable  vine residual dependence structure, but poorly 
for the 1-truncated  regular vine structure. The quite contrary (or complimentary) results are seen for the polychoric algorithm.   The polychoric selection algorithm rather performs extremely well in selecting the true edges in the 1-truncated regular vine residual dependence structure. It is  most accurate for  the initial edges, while it is less accurate for  the final edges. This is because the dependence strength is represented in descending order as $\tau=\{0.40,\ldots,0.10 \}$, so the polychoric selection algorithm is highly reliable to select the edges with stronger dependence. The edges with weaker dependence are not easily quantified and can be approximated with other edges that lead to a similar correlation matrix or even accounted for by the previous trees (factor copula models).

\section{Application}\label{sec:Applications}
In this section we illustrate the proposed methodology by analysing 
$d=20$  items from a subsample of $n=221$ veterans who reported clinically significant  Post Traumatic Stress Disorder  (PTSD) symptoms \citep{Armour-etal2017}. 
The 
 items are divided into four domains: (1) intrusions (e.g.,  repeated, disturbing and unwanted memories), (2) avoidance (e.g., avoiding external reminder of the stressful experience), (3) cognition and mood alterations  (e.g., trouble remembering important parts of the stressful experience) and (4) reactivity alterations (e.g., taking too many risks or doing things that could cause you harm). Each item is answered in a five-point ordinal scale:  ``0 = Not at all", ``1 = A little bit", ``2 = Moderately", ``3 = Quite a bit" and ``4 = Extremely". The dataset and its complete description can be found in \cite{Armour-etal2017} or in the \proglang{R} package \pkg{BGGM} \citep{BGGM_RPackage}.

\begin{table}[!h]
  \centering
    \small
    \caption{Average observed polychoric correlations and semi-correlations for all pairs  of items for the Post Traumatic Stress Disorder dataset, along with the corresponding theoretical semi-correlations for BVN, $t_2$, $t_5$, Frank, Gumbel , and survival Gumbel (s.Gumbel) copulas.} 
      \setlength{\tabcolsep}{49pt}  
    \begin{tabular}{lccc}
    \toprule
      & $\rho_N$ & $\rho_N^-$ & $\rho_N^+$ \\
    \midrule
    Observed  & 0.35 & 0.26 & 0.47 \\
    BVN & 0.35 & 0.16 & 0.16 \\
    $t_2$ & 0.35 & 0.49 & 0.49 \\
    $t_5$ & 0.35 & 0.35 & 0.35 \\
    Frank & 0.35 & 0.10 & 0.10 \\
    Gumbel & 0.35 & 0.11 & 0.37 \\
    s.Gumbel & 0.35 & 0.37 & 0.11 \\
    \bottomrule
    \end{tabular}
  \label{tab:PTSDsemicorr}
\end{table}

For some items, it is plausible that a veteran might be thinking
about the maximum trauma (or a high quantile) of many past events. For example, for the items  in the first domain,
a participant might reflect on past relevant events where an intrusion affected their
life; then by considering the worst case, i.e., the event where the negative effect of an intrusion
 in their life was substantial, they choose an appropriate ordinal response. For some
of the other items, one might consider a median or less extreme harm of past relevant events. To
sum up, the items appear to be a mixed selection between discretized averages and maxima so that a factor
model with more probability in the joint upper tail might be an improvement over a factor model
based on a discretized MVN.

The interpretations as above suggest  that a factor tree  with a combination of Gumbel and BVN or $t_\nu$  copulas might provide a better fit. 
 To further explore the above interpretations, we calculate the average of lower and upper polychoric semi-correlations \citep{Kadhem&Nikoloulopoulos2021,Kadhem&Nikoloulopoulos2021-BJMSP} for all variables to check if there is any overall tail asymmetry.  For comparison, we also report  the theoretical semi-correlations under different choices of copulas. 
Table \ref{tab:PTSDsemicorr} shows averages of the polychoric semi-correlations for all pairs along with the theoretical semi-correlations under different choices of copulas. Overall, we see  that there is more correlation in the joint upper tail than the joint lower tail, suggesting that factor tree copula models with Gumbel  bivariate copulas might be plausible.

We then  select a suitable vine tree structure using the  partial and polychoric selection algorithms proposed in Section \ref{sec:vinetree} and  compute various  discrepancy measures 
between the observed polychoric correlation matrix $\R_{\mathrm{observed}}$ and the correlation matrix $\R_{\mathrm{model}}$ based on factor tree copula models with BVN copulas. We report  the maximum absolute correlation
difference $D_1=\max|\R_{\mathrm{model}} - \R_{\mathrm{observed}}|$, the average absolute correlation
difference $D_2=\mathrm{avg}| \R_{\mathrm{model}} - \R_{\mathrm{observed}}|$ and  the correlation matrix discrepancy measure $D_3=\log\bigl( \det(\R_{\mathrm{model}}) \bigr) - \log\bigl( \det(\R_{\mathrm{observed}})\bigr) + \tr( \R^{-1}_{\mathrm{model}} \R_{\mathrm{observed}} ) - d$. For a baseline comparison, we also compute the discrepancy measures for the 1- and 2-factor copula models with BVN copulas. We aim to obtain a dependence structure that results in the lowest discrepancy measure; this will indicate a  suitable vine structure for the item response data on hand.

After finding a suitable vine structure, 
we 
construct a plausible factor tree copula model, to analyse any type of items, by using the proposed heuristic algorithm in Section  \ref{sec:SelectingCopulas}. 
We use the AIC  at the IFM estimates as a rough diagnostic measure for model selection between the models.  In addition, we  use the  \cite{Vuong1989-Econometrica} procedure that is based on the sample version of the difference in Kullback-Leibler divergence. Let Model 1 and Model 2 have parametric pmfs $\pi^{(1)}_d(\y;\widehat\thbf_1)$ and $\pi^{(2)}_d(\y;\widehat\thbf_1)$, respectively;   $\widehat\thbf_1,\widehat\thbf_2$ are the IFM estimates. The procedure computes the average $\bar D$ of the log differences $D_i=\log\left[\frac{\pi^{(2)}_d(\y_i;\widehat\thbf_2)}{\pi^{(1)}_d(\y_i;\widehat\thbf_1)}\right]$ between the two parametric models.
 \cite{Vuong1989-Econometrica} has shown that asymptotically  
$\sqrt{n}\bar D/s\sim N(0,1)$; $s^2=\frac{1}{n-1}\sum_{i=1}^n(D_i-\bar D)^2$. 
Hence, 
the AIC adjusted Vuong's 95\% CI is $\bar{D}  - n^{-1} [\dim(\widehat\thbf_2) -  \dim(\widehat\thbf_1)] \pm 1.96 \times \frac{1}{\sqrt{n}} \sigma$.  
If it includes 0, then  Model 1 and Model 2 are considered to be non-significantly different, while if it is above 0, then Model 2 is favourable and considered to fit better than Model 1. We will   compare the (1) selected  factor (tree) copula models (Model 2) versus their Gaussian analogues (Model 1), (2)  selected factor tree copula model according to AIC (Model 

\begin{sidewaystable}[htbp]
\begin{minipage}{\textwidth}
  \centering
    \small
  \caption{Measures of discrepancy between the sample and the resulting correlation matrix from the 1-factor, 2-factor, 1-factor tree, and 2-factor tree copula models with BVN copulas for the Post Traumatic Stress Disorder dataset, along with the AICs, Vuong's 95\% CIs,
   for the 1-factor,  2-factor, 1-factor tree, and 2-factor tree copula models with BVN and  selected copulas. Alg.1: partial selection algorithm; Alg.2: polychoric selection algorithm.\label{tab:PTSDapp}}

    \setlength{\tabcolsep}{15.28pt}  

    \begin{tabular}{lccccccccccc}
    \toprule
      & \multicolumn{3}{c}{Factor copula} &   & \multicolumn{3}{c}{1-factor tree copula} &   & \multicolumn{3}{c}{2-factor tree copula} \\
\cmidrule{2-4}\cmidrule{6-8}\cmidrule{10-12}      & 1-factor &   & 2-factor &   & Alg.1 &   & Alg.2 &   & Alg.1 &   & Alg.2 \\
    \midrule
    \multicolumn{12}{l}{BVN copulas} \\
    \midrule
    $D_1$ & 0.40 &   & 0.30 &   & 0.23 &   & 0.20 &   & 0.15 &   & 0.20 \\
    $D_2$ & 0.08 &   & 0.05 &   & 0.05 &   & 0.05 &   & 0.03 &   & 0.05 \\
    $D_3$ & 4.53 &   & 2.80 &   & 1.75 &   & 1.83 &   & 1.17 &   & 1.75 \\
    \#parameters & 20 &   & 39 &   & 39 &   & 39 &   & 58 &   & 58 \\
    AIC & 12031.1 &   & 11764.0 &   & 11632.4 &   & 11642.1 &   & 11549.1 &   & 11611.8 \\
    \midrule
    \multicolumn{12}{l}{Selected copulas} \\
    \midrule
    \#parameters & 20 &   & 40 &   & 39 &   & 39 &   & 59 &   & 59 \\
    AIC & 11800.4 &   & 11413.5 &   & 11355.3 &   & 11344.89 &   & 11189.1 &   & 11240.3 \\
    Vuong's 95\% CI\footnote{Selected factor (tree) copula models versus their Gaussian analogues. } & ( 0.21, 0.63) &   & (0.25, 0.79) &   & (0.37, 0.89) &   & (0.43, 0.91) &   & ( 0.54, 1.09) &   & (0.58, 1.11) \\
    Vuong's 95\% CI\footnote{Selected 2-factor tree copula model with Alg.1 versus other fitted models with BVN copulas.} & (1.50, 2.31)  &   & (0.99, 1.67)  &   & (0.79, 1.40) &   & (0.83, 1.40) &   & - &   & (0.69, 1.24) \\
    Vuong's 95\% CI\footnote{Selected 2-factor tree copula model with Alg.1 versus other fitted models with selected copulas.} & (1.17, 1.80)  &   & (0.60, 1.02)  &   & (0.30, 0.63)  &   & (0.27, 0.61) &   & - &   & (-0.002, 0.23) \\
    \bottomrule
    \end{tabular}%

    \end{minipage}
\end{sidewaystable}

\noindent 2) versus all the other factor (tree) copula models with BVN copulas (Model 1), and (3) selected factor tree copula model according to AIC (Model 2) versus all the other factor (tree) copulas models with selected copulas (Model 1).
Note in passing that the 2-factor  (tree) copula models with BVN copulas will have one dependence  parameter less 
as one copula in the second factor is set to independence for identification purposes.

Table  \ref{tab:PTSDapp} shows that the sample correlation matrix of the data has a 2-factor tree  structure according to the discrepancy measures. The table also gives the AICs and  the 95\% CIs of Vuong's tests for all the fitted models. The best fitted model, based on AIC values, is the 2-factor tree copula model obtained from the partial selection algorithm. The best fitted 2-factor tree copula model has the $t_2$ for the 1st tree, Gumbel for the 2nd tree, and $t_5$ for the 3rd tree.  From the Vuong's 95\% Cls  it is shown that 2-factor tree copula model provides a big improvement over its Gaussian analogue and  outperforms all the other fitted models except  the 2-factor tree obtained from the polychoric selection algorithm.  
The tree selection algorithms might not yield into the same `true' vine tree, however, closely approximated factor tree copula models are achieved.
The factor tree copula model is mostly constructed with $t_2$ bivariate copulas which are suitable for both positive and negative dependence, however  the highest dependence is found in the 2nd factor which is constructed with Gumbel copulas. This is in line with both the initial interpretations and preliminary analysis which suggest that some items can be  considered as discretized maxima.

\begin{table}[!h]
  \centering
    \small
  \caption{ Estimated copula parameters and their standard errors (SE) in Kendall's $\tau$ scale for the selected 2-factor and  2-factor tree copula models obtained from the partial selection algorithm for the Post Traumatic Stress Disorder dataset. 
\label{tab:PTSDappEST}}
  
      \setlength{\tabcolsep}{7.63pt}  
    \begin{tabular}{cccccccccccccccc}
    \toprule
      & \multicolumn{5}{c}{2-factor copula} &   & \multicolumn{9}{c}{2-factor tree copula} \\
\cmidrule{2-6}\cmidrule{8-16}    Tree & \multicolumn{2}{c}{1st factor} &   & \multicolumn{2}{c}{2nd factor} &   & \multicolumn{2}{c}{1st factor} &   & \multicolumn{2}{c}{2nd factor} &   & \multicolumn{3}{c}{Vine model} \\
\cmidrule{2-3}\cmidrule{5-6}\cmidrule{8-9}\cmidrule{11-12}\cmidrule{14-16}    Copula & \multicolumn{2}{c}{$t_2$} &   & \multicolumn{2}{c}{Gumbel} &   & \multicolumn{2}{c}{$t_2$} &   & \multicolumn{2}{c}{Gumbel} &   & \multicolumn{3}{c}{ $t_5$} \\
    \midrule
    Items & $\hat\tau$ & SE &   & $\hat\tau$ & SE &   & $\hat\tau$ & SE &   & $\hat\tau$ & SE &   & $ \mathcal{E}$ & $\hat\tau$ & SE \\
    \midrule
    1 & 0.16 & 0.06 &   & 0.49 & 0.04 &   & -0.17 & 0.06 &   & 0.50 & 0.04 &   & $1,18$ & -0.18 & 0.06 \\
    2 & 0.11 & 0.06 &   & 0.49 & 0.04 &   & -0.08 & 0.06 &   & 0.45 & 0.04 &   & $18,17$ & 0.22 & 0.06 \\
    3 & 0.14 & 0.06 &   & 0.54 & 0.04 &   & -0.12 & 0.06 &   & 0.52 & 0.04 &   & $18,14$ & -0.20 & 0.07 \\
    4 & 0.32 & 0.06 &   & 0.56 & 0.05 &   & -0.34 & 0.06 &   & 0.57 & 0.05 &   & $18,10$ & -0.10 & 0.06 \\
    5 & 0.21 & 0.06 &   & 0.55 & 0.04 &   & -0.21 & 0.06 &   & 0.56 & 0.04 &   & $10,11$ & 0.36 & 0.05 \\
    6 & 0.13 & 0.06 &   & 0.28 & 0.05 &   & -0.13 & 0.06 &   & 0.26 & 0.05 &   & $11,9$ & 0.29 & 0.06 \\
    7 & 0.11 & 0.06 &   & 0.40 & 0.04 &   & -0.09 & 0.06 &   & 0.39 & 0.04 &   & $9,2$ & -0.18 & 0.06 \\
    8 & -0.03 & 0.06 &   & 0.21 & 0.05 &   & 0.04 & 0.06 &   & 0.19 & 0.05 &   & $2,3$ & 0.26 & 0.06 \\
    9 & -0.17 & 0.06 &   & 0.38 & 0.04 &   & 0.24 & 0.06 &   & 0.33 & 0.04 &   & $3,20$ & 0.05 & 0.07 \\
    10 & 0.16 & 0.06 &   & 0.34 & 0.05 &   & -0.12 & 0.06 &   & 0.30 & 0.04 &   & $2,16$ & 0.13 & 0.06 \\
    11 & 0.09 & 0.06 &   & 0.52 & 0.04 &   & -0.07 & 0.06 &   & 0.48 & 0.04 &   & $16,15$ & 0.17 & 0.06 \\
    12 & -0.23 & 0.06 &   & 0.50 & 0.04 &   & 0.28 & 0.06 &   & 0.50 & 0.04 &   & $9,4$ & 0.29 & 0.08 \\
    13 & -0.35 & 0.06 &   & 0.55 & 0.05 &   & 0.34 & 0.05 &   & 0.49 & 0.05 &   & $20,5$ & 0.05 & 0.07 \\
    14 & -0.37 & 0.05 &   & 0.41 & 0.05 &   & 0.35 & 0.05 &   & 0.36 & 0.05 &   & $14,13$ & 0.27 & 0.07 \\
    15 & -0.09 & 0.06 &   & 0.48 & 0.04 &   & 0.11 & 0.06 &   & 0.44 & 0.04 &   & $5,6$ & 0.12 & 0.07 \\
    16 & -0.08 & 0.06 &   & 0.31 & 0.05 &   & 0.10 & 0.06 &   & 0.28 & 0.04 &   & $6,7$ & 0.23 & 0.06 \\
    17 & -0.04 & 0.06 &   & 0.34 & 0.04 &   & 0.04 & 0.06 &   & 0.33 & 0.04 &   & $7,19$ & -0.21 & 0.06 \\
    18 & -0.06 & 0.06 &   & 0.45 & 0.04 &   & 0.12 & 0.06 &   & 0.46 & 0.04 &   & $16,8$ & 0.12 & 0.06 \\
    19 & -0.26 & 0.06 &   & 0.45 & 0.04 &   & 0.28 & 0.06 &   & 0.43 & 0.04 &   & $19,12$ & 0.08 & 0.07 \\
    20 & -0.11 & 0.06 &   & 0.41 & 0.04 &   & 0.13 & 0.06 &   & 0.40 & 0.04 &   & -  &  - &  -\\
    \bottomrule
    \end{tabular}
\end{table}

Table \ref{tab:PTSDappEST} includes the copula parameter estimates in Kendall's $\tau$ scale and their standard errors (SE) for the selected 2-factor and 2-factor tree copula models. The latter is obtained from the partial selection algorithm. To make it easier to compare strengths of dependence, 
we convert the BVN/$t_\nu$ and  
 Gumbel/s.Gumbel 
 copula parameters to Kendall's $\tau$'s 
 via  the relation $\tau(\theta)=\frac{2}{\pi}\arcsin(\th)$  and   (\ref{tauGumbel}), respectively.  Interestingly, the Kendall's $\tau$'s in the 2-factor copula model are roughly equivalent to the estimates in the 1st and 2nd factors of the 2-factor tree copula model. Most of the dependence is captured in the first two trees, resulting in weak to medium residual dependencies in the 1-truncated vine copula model, but significantly larger from independence. Overall, the items, in the Markov tree, are mostly positively associated to one another with only few negative conditional dependencies.  The residual dependencies reveal that there is stronger association between the 10th and 11th items that are ``Blame of self or others" and ``Negative trauma-related emotions", respectively. 
In addition, there is moderate association between  items 9 and 11 that are ``Negative beliefs" and ``Negative trauma-related emotions", respectively. With similar moderate dependence found between items 9 and 4 that are ``Negative beliefs" and ``Emotional cue reactivity", respectively.

\section{Discussion}\label{sec:Discussion}
We have proposed combined factor/truncated vine copula models to capture the residual dependence for item response data. They form conditional dependence of the items given the latent variables, and go beyond the  factor models where the items are conditionally independent given the latent variables.
By combining the factor copula models with an 1-truncated vine copula model, we  construct conditional dependence models given very few interpretable latent variables. The combined factor/truncated vine structure has the form of (i) primary dependence being
explained by one or more latent variables, and (ii) conditional dependence of item response variables
given the latent variables \citep{Joe2018}. They are especially useful and interpretable
  when there are a few  latent variables that can explain most but not all of the dependence in the item responses.

The flexibility of the factor tree  copula  models endorses the significance of model selection. In practice, one has to first select the 1-truncated vine tree structure  $ \mathcal{E}$ and then suitable bivariate copulas to account for more probability in the one or both joint tails. We tackle these model selection issues by proposing heuristic algorithms to choose a plausible factor tree copula model that can adequately capture the (residual) dependencies among the item responses. 
We have shown that the proposed models provide a substantial improvement over the  1-factor and 2-factor  (tree) copula models with selected (BVN) copulas on the basis of the AIC and Vuong's statistics. The 1-factor and 2-factor  tree copula models with BVN can be
viewed as first order models if models based on other tail dependent copulas are called. We consider the 1-factor and 2-factor tree copula models to be reasonable parsimonious models as most of the dependence is explained via the first few trees in the factor model. 
This is because that for all the bivariate margins to have upper/lower tail dependence, it only suffices that the bivariate copulas in the first trees (factor part) to have upper/lower tail dependence and is not necessary for the bivariate copulas in the higher trees after the 1-truncated vine to have tail dependence  \citep{Joe&Li&Nikoloulopoulos2010}.

The proposed models are reproducible as the conditional independence and residual dependence parts  are modelled separately. 
The residual dependencies are  taken into account by a Markov tree without changing anything to the conditional independence  model part.  The use of a Markov tree for the residual dependence is a new direction for parsimonious
dependence.
 This reproducibility  as per the terminology in  \cite{liang02}, means that we can remain within a well-known and conceptually attractive framework as offered by the factor copula models when applying a factor tree copula model. This  will be attractive  to practitioners that have a basic and conceptual understanding of factor models, but are less familiar with complicated models that are available to approach the problem of residual dependence. The main change in the factor copula model is only in the formulation of the joint conditional distribution, while the  conditional part of the model, i.e., the unique loading parameters, these are $\hat\tau$s converted to normal copula parameters $\hat \th_{1j}$ and $\hat\th_{2j}$ and  then to loadings with the relations in Section \ref{sec:vinetree}, is left intact.

\section*{Software}
\proglang{R} functions for estimation, simulation and model selection of the factor tree copula models will be part of the next major release of the \proglang{R} package \pkg{FactorCopula} \citep{Kadhem&Nikoloulopoulos-2020-package}. 

\section*{Acknowledgements}
 The simulations presented in this paper were carried out on the High Performance Computing Cluster supported by the Research and Specialist Computing Support service at the University of East Anglia.


\end{document}